\documentclass[10pt,journal]{IEEEtran}

\usepackage{bm} 

\usepackage{epsfig}
\usepackage{amssymb} 
\usepackage[tbtags]{amsmath} 

\usepackage{latexsym}
\usepackage{euscript}
\usepackage{graphics,eepic,epic,psfrag}
\usepackage{enumitem}
\usepackage{xhfill}
\usepackage{relsize}
\usepackage{exscale}
\usepackage{stfloats}%

\usepackage[font = footnotesize]{caption}
\usepackage{subcaption}
\usepackage{graphicx}
\usepackage{cite}  
\usepackage{xpatch}
\usepackage{algorithmicx}
\usepackage{romannum}

\usepackage{comment}

\usepackage{algpseudocode}
\usepackage[linesnumbered,ruled,vlined]{algorithm2e}

\usepackage{stfloats}
\usepackage{url}

\DeclareMathOperator*{\argmin}{arg\,min}

\newcommand\numberthis{\addtocounter{equation}{1}\tag{\theequation}}
\newcommand{\upperRomannumeral}[1]{\uppercase\expandafter{\romannumeral#1}}

\newlength\myindent
\begin{document}
\pagenumbering{arabic}

%
\title{Learning Beamforming Codebooks for Active Sensing with Reconfigurable Intelligent Surface}

\author{Zhongze~Zhang,~\IEEEmembership{Graduate Student Member,~IEEE,}
        and~Wei~Yu,~\IEEEmembership{Fellow,~IEEE}
\thanks{Manuscript submitted to \emph{IEEE Transactions on Wireless Communications} on September 6, 2024, revised on January 12, 2025, accepted on March 5, 2025. 
Zhongze Zhang is with The Edward S. Rogers Sr. Department
of Electrical and Computer Engineering, University of Toronto, Toronto,
ON M5S3G4, Canada, and the Department of Electrical and Electronic Engineering, University of Melbourne, Victoria, Australia (E-mails: ufo.zhang@mail.utoronto.ca). 
Wei Yu is with The Edward S. Rogers Sr. Department
of Electrical and Computer Engineering, University of Toronto, Toronto,
ON M5S3G4, Canada (E-mails: weiyu@comm.utoronto.ca).  
This work is supported by the Natural Sciences and Engineering Research Council of Canada via the Canada Research Chairs program. The materials in this paper have been accepted in part at the IEEE Workshop on Signal Processing Advances in Wireless Communications (SPAWC), Lucca, Italy, September 2024\cite{codebookspawc}.
}
}

\maketitle

\begin{abstract} 

This paper explores the design of beamforming codebooks for the base station (BS) and for the reconfigurable intelligent surfaces (RISs) in an active sensing scheme for uplink localization, in which the mobile user transmits a sequence of pilots to the BS through reflection at the RISs, and the BS and the RISs are adaptively configured by carefully choosing BS beamforming codeword and RIS codewords from their respective codebooks in a sequential manner to progressively focus onto the user. Most existing codebook designs for RIS are not tailored for active sensing, by which we mean the choice of the next codeword should depend on the measurements made so far, and the sequence of codewords should dynamically focus reflection toward the user. Moreover, most existing codeword selection methods rely on exhaustive search in beam training to identify the codeword with the highest signal-to-noise ratio (SNR), thus incurring substantial pilot overhead as the size of the codebook scales. This paper proposes a learning-based approach for codebook construction and for codeword selection for active sensing. The proposed learning approach aims to locate a target in the service area by recursively selecting a sequence of BS beamforming codewords and RIS codewords from the respective codebooks as more measurements become available without exhaustive beam training. The codebook design and the codeword selection fuse key ideas from the vector quantized variational autoencoder (VQ-VAE) and the long short-term memory (LSTM) network to learn respectively the discrete function space of the codebook and the temporal dependencies between measurements.

\end{abstract}


%
\IEEEpeerreviewmaketitle

\section{Introduction}

\IEEEPARstart{R}{}econfigurable intelligent surface (RIS) is a planar surface made up of numerous passive elements, each capable of adjusting the phase of an incoming electromagnetic wave with minimal power consumption \cite{bible, survey_ris}.
The device is typically placed in the reflecting path between the transceivers, with its configuration wirelessly controlled by the transceivers via a control link. 
However, the control
link typically has a limited capacity, so a straightforward RIS
control protocol, which sends the settings of continuous phase shifts of
the RIS elements through the control link\cite{hguo, nulling, scheduling}, is often infeasible. 
Codebook-based limited control link rate protocol can substantially reduce the control overhead \cite{codewordsurvey, randomcodebook}. With the RIS codebook stored at the controller and at the RIS, the controller only needs to send the codeword index in order to configure the RIS. 
Substantial research progress has been made to study the design of such a codebook
\cite{codebook_euclideanmax,codebook_euclideanmax2, fieldtest, ringtype,env_aware, codebook_learning_RL, codebook_RL_3, codebook_learning_DNN, hejiguang, hierarchical1,hiarchicalwuqq, hierarchical2layer}.





This paper considers the design of an RIS codebook and a base station (BS) beamforming codebook that enables active sensing
in an uplink localization setting. By active sensing, we envision a setting
in which a mobile user repeatedly transmits pilot symbols; the
BS receives the pilots through reflections at possibly multiple RISs; the reflection coefficients at the RISs and the BS beamforming vector are actively
reconfigured by selecting RIS codewords from the RIS codebook and BS beamforming codewords from the BS beamforming codebook as a function of
existing measurements made so far, in order to eventually determine the location of the user
\cite{active2023icc,twcactive}.
Such a sequential sensing strategy makes use of the wealth of knowledge contained in existing measurements to recursively select the sequence of codewords from respective codebooks to 
enable the RISs and the BS to focus onto the user progressively over time as more measurements become available.
We aim to answer two questions: \romannum{1}) \emph{Codebook construction}: how to construct codebooks that enable active sensing? \romannum{2}) \emph{Codeword selection}: Given a codebook, how to select the codeword for the next pilot based on the existing measurements?

Active sensing is a new paradigm that can significantly improve the communication objective \cite{sohrabi2021active} and sensing accuracy\cite{twcactive}, but the design of codebooks that support such a paradigm is challenging, because designing an optimal codebook requires searching through a complex discrete functional landscape. 
Among the works that are most related to the concept of active sensing is the hierarchical codebook based approach \cite{hierarchical1,hejiguang, hiarchicalwuqq,hierarchical2layer}, where the RIS codebook is constructed based on the channel model, and the sequence of RIS codewords is selected using sectional search that gradually narrows the search range. 
However, this approach is sensitive to noise, and exhaustive search is required within the narrowed search range in order to identify the optimal codeword. This is a process in which the optimal RIS codeword is found by exhaustive beam training using every candidate codeword in the narrowed range, i.e., by measuring the resulting received power and selecting the codeword with the largest signal-to-noise ratio (SNR). In essence, the RIS probes the search area using different beams along multiple directions. 
This takes up substantial beam training time as the network scales. 
Moreover, 
such a hierarchical codebook based method is greedy in the sense that it only empowers exploitation (i.e., focusing in directions of interest) at the expense of exploration (i.e., probing different directions) in the beam space. Achieving a proper balance between exploitation and exploration is crucial in harnessing the full potential of active sensing.

This paper proposes a data-driven approach for constructing an RIS codebook and a BS beamforming codebook, and based on the learned codebooks, for recursively selecting a sequence of RIS codewords and BS beamforming codewords from the respective codebooks based on the measurements received so far for localization, without exhaustive beam training. The conventional paradigm often treats the \emph{codebook construction} and \emph{codeword selection} problems separately. Here, the proposed neural network learns the solutions to the two problems together. The codebook is learned through the codeword selection process, where a subset of the codewords in the codebook are updated one training sequence at a time until all codewords are fully trained via \emph{forward propagation} and \emph{backward propagation}. (Here, a training sequence consists of a sequence of pilots from the user.)


Conceptually, the deep learning algorithm works as follows. Given a codebook of a fixed number of trainable codewords, 
during \emph{forward propagation}, the learning algorithm recursively receives new measurements from a user and maps the new and historical measurements to a codeword, which is to be used to make the next measurement. For a sensing sequence involving $T$ measurements, $T$ codewords are selected to survey the environment. During \emph{backward propagation}, the sensing loss is computed, and the selected codewords are updated 
to reduce training loss. 
Through many training sequences with users at different locations, many sets of $T$ codewords (which can be overlapping) are selected and updated until all codewords are updated and they reach an equilibrium. Given the optimized codebook, the learning algorithm can automatically draw a sequence of codewords based on the recurring measurements without exhaustive beam training to locate any user within the service area. 

This proposed learning-based approach is based on a combination of vector quantized variational autoencoder (VQ-VAE)\cite{discretereplearning} and long short-term memory (LSTM) network \cite{LSTM} to respectively learn the discrete function space of the codebook and learn the temporal relationship between different measurements over a long period of time. Specifically, we make use of LSTM units as a history compressor to construct an information vector of fixed dimension from existing measurements and to extract temporal dependencies from a sequence of inputs. 
However, the use of LSTM alone is insufficient as the mapping from information vector to codeword index is challenging for the neural network to learn, because such mappings are onto a discrete set. Hence, we make use of the idea of VQ-VAE to map the learned information vector to the discrete codebook space via vector quantization (VQ), and perform gradient estimation to approximate the missing gradient of VQ operation to enable the training of the neural network. 
This is a remodel of the neural network architecture introduced in the previous work \cite{twcactive}, which tackles continuous optimization of active sensing vector in a codebook-free setting using LSTM alone, to a distinct solution tailored for a class of discrete optimization problems of codebook design for active sensing tasks. 
Numerical results show that the proposed algorithm can effectively learn an RIS codebook and a BS beamforming codebook that enable active sensing, i.e., a sequence of RIS codewords and BS beamforming codewords can be adaptively chosen from the respective RIS codebook and the BS beamforming codebook without exhaustive beam training to enable accurate localization. The resulting learned codebooks contain interpretable beampatterns. 

We highlight a related effort in automatic speech recognition, where VQ-VAE is integrated with LSTM models \cite{shi2022vqtrnntransducersusing}. In that work, the information vector is quantized using a fixed, predefined codebook. In contrast, the approach of this paper adaptively learns the codebook by quantizing the RIS configuration into codewords, rather than relying on a static design. This learning-based strategy enables a design of the codebook to be tailored to the specific environmental and system parameters.

\subsection{Related Work}
\emph{Codebook construction} for RIS to improve the network performance is of great interest. 
Many works 
design codebooks that contain diverse RIS patterns to generate different reflection channels \cite{codebook_euclideanmax, codebook_euclideanmax2, fieldtest, ringtype}, or are adaptive to the time-varying channel, site-specific environment and hardware characteristics\cite{env_aware,codebook_learning_RL, codebook_learning_DNN, codebook_RL_3}. The authors of \cite{codebook_euclideanmax} and \cite{codebook_euclideanmax2} propose RIS codebooks with discrete and continuous phase shifts respectively to improve the received SNR at the receiver; the codewords are Euclidean-distance maximized to ensure a diverse set of reflection channels. In \cite{fieldtest}, Discrete Fourier Transform (DFT) codebook is used 
with each codeword representing the reflection of an incident beam in a certain beam direction. 
The authors of \cite{ringtype} propose a ring-type codebook based on the Fresnel principle to achieve SNR enhancement and mirror grating lobe impression in the near-field region. 
Advocating that the codebook should be designed according to channel statistics, 
the authors of \cite{env_aware} use statistical channel state information (CSI) to design an RIS codebook that better accounts for the underlying wireless channel. 
Further, the authors of \cite{codebook_RL_3, codebook_learning_RL} and \cite{codebook_learning_DNN} use reinforcement learning to design RIS codebooks that are adaptive to site-specific environments, time-varying channels and hardware characteristics. 
All of these works emphasize the importance of a carefully designed RIS codebook and the resulting improvement in network performance. 
However, these existing codebooks are not designed for active sensing to enable sequential drawing of codewords to gradually focus toward the user. 

We also remark that most existing works perform \emph{codeword selection} via exhaustive search\cite{codebook_euclideanmax, codebook_euclideanmax2, fieldtest,env_aware,codebook_learning_RL,codebook_learning_DNN}, where the optimal RIS codeword is found by exhaustive beam training using every codeword in the codebook and by measuring the resulting received power and selecting the codeword with the largest SNR. 
Doing so requires substantial beam training time for the RIS to probe the search area using different beams along multiple directions. 
Though efforts are made to reduce the size of the codebook \cite{codebook_learning_RL}, substantial beam training overhead is inevitable as the network scales.

%

Recently, active sensing has shown promising performance in reducing beam training time for a number of network applications, including mmWave initial beam alignment problem\cite{sohrabi2021active, GRU , pingpong}, precoder and combiner design problem for reciprocal multi-input multi-output (MIMO) channel\cite{taomimo}, beam tracking problem with RIS \cite{hantracking}, and localization problem with RIS in single-path \cite{twcactive} and multi-path environment \cite{yinghanactive}. The above works utilize recurrent neural network (RNN) based solutions that dynamically design the sensing beamformers based on previously received pilots to gradually focus onto the target. However, the proposed neural networks in the above works cannot be directly generalized to the codebook setting as they only operate in the codebook-free setting with differentiable functional landscape that the neural network can easily navigate. 
In this paper, we focus on the codebook design for active sensing, which entails a challenging discrete function landscape for a neural network to learn. 

In parallel with the codebook-based protocol investigated in this paper, some works explore autoencoder-based frameworks to reduce control link rate, where the encoder at the transceivers compresses the RIS configurations, and the decoder at the RIS reconstructs them using a neural network \cite{compress, lightweight}. These approaches focus on achieving perfect reconstruction, whereas we use the localization performance as the ultimate objective.






\subsection{Main Contribution}

This paper proposes a learning-based codebook design that enables active sensing, where an adaptive sequence of sensing vectors (including multiple RIS reflection coefficients and the BS beamforming vector) can be drawn from the respective RIS codebook and BS beamforming codebook based on the available measurements of the environment. The selected BS beamforming codewords and RIS codewords would gradually focus onto the user and would ultimately lead to accurate localization. This is achieved by a data-driven approach to learn the mapping from existing measurements to the sensing codewords to be used in the next time frame within the discrete codebook space. 
Specifically, we use a combination of LSTM network and VQ-VAE network, for their respective ability to capture the temporal dependencies between inputs over an extended period of time \cite{RNN} and to capture the discrete representations of a latent variable \cite{discretereplearning}. 
At each time frame, an LSTM unit accepts a new measurement of the environment, which is used along with the historical measurements to update the hidden state vector and subsequently output a sensing vector (i.e., the BS beamforming vector and RIS configurations). The LSTM-designed sensing vectors are mapped to codewords in their respective codebook via VQ. Since there is no gradient defined for the VQ operation, gradient approximation is performed to estimate the missing gradient prior to the VQ operation. 
Subsequently, the selected codewords are used to obtain a new measurement in the next time frame. 
In each training sequence (which consists of multiple time frames), we select a series of sensing codewords. Those selected sensing codewords (along with the weights of the LSTM network) are updated in backward propagation to reduce a composite loss function. 
Through many training sequences, many sets of sensing codewords are selected and updated in a similar fashion until all codewords in the codebook are updated to reach an equilibrium. 

The contributions of this paper are as follows: 
\begin{enumerate}
    \item A codebook-based active sensing scheme is proposed to locate the user equipment (UE) in a multi-RIS-assisted multi-input single-output (MISO) network, where the sequence of RIS codewords and BS beamforming codewords are adaptively chosen from the respective codebook as more measurements become available to enable better localization accuracy. 

    \item A codebook construction scheme is proposed to learn the BS beamforming codebook and the RIS codebook that contain a diverse set of codewords with broad and narrow beams to enable adaptive selection of codewords for gradual focusing toward the UE in active sensing. 


    \item A codeword selection strategy is proposed for drawing a sequence of codewords from the codebooks for a localization task without exhaustive beam training, which saves valuable channel coherence time. This is achieved via VQ which can be parallelized without requiring additional pilot training to identify the optimal codeword. 
    
    \item An LSTM architecture incorporating key ideas from VQ-VAE is proposed to learn to summarize state information of available temporal input and to map the information to the discrete codebook space.
    The loss function is carefully designed to train the LSTM network, RIS codebook and BS beamforming codebook jointly. 
    We term the proposed neural network architecture VQ-Codebook (VQ-C).    
    \item The proposed codebook design and codeword selection scheme produces interpretable results, showing that the selected codewords gradually focus toward the UE to improve the received SNR for localization. 
    
\end{enumerate}

Numerical results show that the proposed algorithm can effectively learn a BS beamforming codebook and an RIS codebook to enable active sensing. During the localization process, a sequence of BS beamforming codewords and RIS codewords can be adaptively chosen without exhaustive beam training to perform accurate localization, with performance approaching its codebook-free counterpart. 

The conference version of this journal paper focuses on RIS codebook design for active sensing in a single RIS single-input single-output (SISO) network\cite{codebookspawc}. The current paper extends to a joint RIS codebook and BS beamforming codebook design for active sensing in a multi-RIS MISO network. 
The proposed neural network can also be extended to MIMO networks, where the UE and the BS are equipped with multiple antennas, by jointly optimizing the RIS pattern, BS beamforming vector, and UE beamforming vector (e.g., mapping the LSTM hidden state vector to the UE beamforming vector). Here, we focus on the more challenging SISO and MISO setups to highlight the localization benefits enabled by RIS through active sensing, as accurate localization in SISO network is difficult without RIS. Furthermore, the algorithm can be extended to multi-user localization using time-multiplexing.

\subsection{ Paper Organization and Notations}
The rest of this paper is structured as follows. 
Section \ref{sec.background} describes the system model, the pilot transmission protocol, and the problem formulation under the active sensing framework. 
Section \ref{sec.rnn} introduces the proposed VQ-C architecture. 
Section \ref{sec.numerical_irs_single} and Section \ref{sec.numerical_irs_multi} present the numerical results in single RIS SISO network and multi-RIS MISO network respectively.  
 Section \ref{sec.conclusions} concludes the paper.

We use $y$, $\bm{y}$, and $\bm{Y}$ to denote scalar, vector, and matrix respectively, $(\cdot)^\top$ and $(\cdot)^{\sf H}$ to denote transpose and Hermitian, $|\cdot|$ to denote modulus, $\| \cdot \|_2^2$ to denote squared norm, 
$[\bm{y}]_j$ to denote the $j$-th element of the vector $\bm{y}$, 
$[\bm{Y}]_{i,j}$ to denote the $(i,j)$-th element of the matrix $\bm{Y}$, $[\bm{Y}](:,j)$ to denote the $j$-th column of the matrix $\bm{Y}$, 
and diag($\bm{y}$) to denote a diagonal matrix with entries of $\bm{y}$ on the diagonal. 
We represent complex Gaussian distributions with $\mathcal{C}\mathcal{N}(\cdot,\cdot)$; the real and imaginary parts of a complex value with $\mathcal{R}(\cdot)$ and $\mathcal{I}(\cdot)$, respectively; and the expectation of a random variable with $\mathbb{E}(\cdot)$. The floor function is denoted by $\lfloor \cdot \rfloor$, and the modulo operation by ${\rm mod}(\cdot)$.

\section{Active Sensing via Beamforming Codebooks}
\label{sec.background}

\subsection{System Model}

\begin{figure}[t]
\centering
\includegraphics[width=\columnwidth]{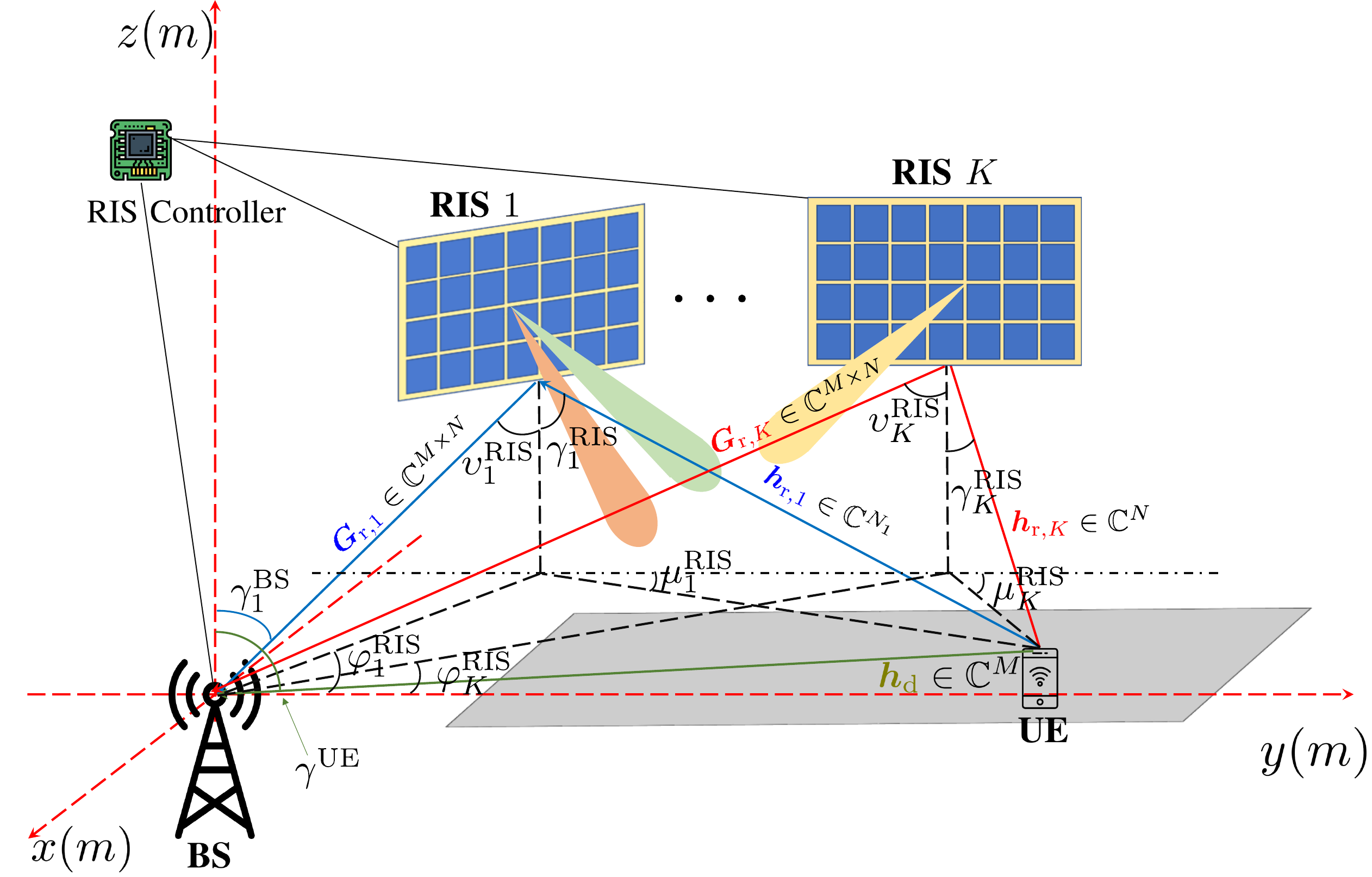}%
\caption{Multi-RIS MISO network.}%
\label{fig.setup_intro}%
\end{figure}

Consider a localization problem in an RIS-assisted MISO network, in which a BS equipped with multiple antennas but limited RF chain seeks to locate a single-antenna UE with the help of $K$ planar RISs. 
The BS and RISs are placed as in Fig.~\ref{fig.setup_intro} to localize the potential user in the area. 
We use $\bm{p}^{\rm BS}$ and $\bm{p}_k^{\rm RIS}$ to denote the known positions of the BS and the $k$-th RIS respectively.
The unknown UE position is denoted as $\bm{p} = [x,y,z]^\top$. This is the same localization problem as already considered in \cite{twcactive} in the codebook-free setting.

The reflection coefficients of the RISs are controlled by an RIS controller. Due to the limited capacity of the control link, we adopt a codebook-based control protocol. With the codebook stored at the BS and the RISs, only the codeword indices need to be transmitted. Let $\bm{\Theta}^{\rm RIS} \in \mathbb{C}^{N\times V}$ denote the RIS codebook shared amongst the $K$ RISs. The codebook contains $V$ codewords, each of which takes the form
\begin{equation}
    \bm{\theta} = [e^{j\delta_{1}},  e^{j\delta_{2}}, \cdots , e^{j\delta_{N}}]^\top\in\mathbb{C}^{N},
\end{equation}
where $N$ denotes the number of reflection coefficients at each RIS with $\delta_{n} \in [0,2\pi)$ as the phase shift of the $n$-th element. Each element satisfies the constant modulus constraint i.e.,  $\left|[\bm{\theta}]_n\right| = 1$.

In this paper, we assume that the BS is equipped with $M$ antennas but only with one radio frequency (RF) chain, so it can only sense the channel through an analog beamforming vector. In addition, the analog beamformers must consist only of phase shifters. We also use a codebook-based approach for BS beamforming and use $\bm{W}^{\rm BS} \in \mathbb{C}^{M\times B}$ to denote the analog BS beamforming codebook. The codebook contains $B$ codewords, each of which takes the form
\begin{equation}
    \bm{w} = [e^{j\omega_{1}},  e^{j\omega_{2}}, \cdots , e^{j\omega_{M}}]^\top\in\mathbb{C}^{M},
\end{equation} 
whose entries also satisfy a constant modulus constraint, i.e., 
$\omega_m \in [0,2\pi)$ and $\left| [\bm{w}]_m\right| = 1$.


A \emph{block-fading channel model} is adopted in which the channel coefficients are assumed to remain constant over multiple time frames during a coherence block, but change independently from block to block. 
We consider an indoor scenario with low UE mobility, where the coherence time exceeds the sensing period. 
As shown in Fig.~\ref{fig.setup_intro}, we use $\bm{h}_{\rm{d}}\in\mathbb{C}^M$ to denote the direct channel between the BS and the UE, $\bm{h}_{{\rm{r}},k}\in\mathbb{C}^{N}$ to denote the reflection channel from the UE to the $k$-th RIS and $\bm{G}_{{\rm{r}},k} \in\mathbb{C}^{M \times N }$ to denote the channel from the $k$-th RIS to the BS.
Multiple reflections across the multiple RISs are ignored because of the considerable path loss associated with these reflections.
The direct channel and reflection channels are assumed to follow Rician fading model as in 
\begin{subequations} 
\begin{align}
\bm{h}_{\rm d}  =\;& \rho\left(\sqrt{{\epsilon}/{(1+\epsilon)}} \bm{\tilde{h}}_{{\rm d}}^{\rm LOS} + \sqrt{{1}/{(1+\epsilon)}}\bm{\tilde{h}}_{{\rm d}}^{\rm NLOS}\right)\label{hdrician},\\
\bm{h}_{{\rm r},k} =\; & \kappa_k\left(\sqrt{{\epsilon}/(1+\epsilon)} \bm{\tilde{h}}_{{\rm r},k}^{\rm LOS} + \sqrt{{1}/{(1+\epsilon)}}\bm{\tilde{h}}_{{\rm r},k}^{\rm NLOS}\right),\\
\bm{G}_{{\rm r},k} =\; & \xi_k\left(\sqrt{{\epsilon}/{(1+\epsilon)}}\bm{\tilde{G}}_{{\rm r},k}^{\rm LOS} + \sqrt{{1}/{(1+\epsilon)}}\bm{\tilde{G}}_{{\rm r},k}^{\rm NLOS}\right).
\end{align}
\end{subequations}




The path loss components of the channels, $\rho$,
$\kappa_k$ and $\xi_k$, contain distance information about the location of UE, 
where $\rho$, $\kappa_k$ and $\xi_k$ denote the path loss between the BS and the UE, between the $k$-th RIS and the UE, and between the $k$-th RIS and the BS respectively.
The line-of-sight (LOS) components of the channels, $\bm{\tilde{h}}_{{\rm r},k}^{\rm LOS}$, $\bm{\tilde{G}}_{{\rm r},k}^{\rm LOS}$ and  $\bm{\tilde{h}}_{{\rm d}}^{\rm LOS}$, contain angular information about the location of UE.
The modelling of the LOS components is detailed in Section \ref{sec.numerical_irs_single}. 
The non-line-of-sight (NLOS) components of the channels, $\bm{\tilde{h}}^{\textrm{NLOS}}_{{\rm d}}$,
$\bm{\tilde{h}^{\textrm{NLOS}}}_{{\rm r},k}$ and $\bm{\tilde{G}}_{{\rm r},k}^{\textrm{NLOS}}$, have entries modelled as i.i.d. standard Gaussian distribution, i.e., $[\bm{\tilde{h}}^{\textrm{NLOS}}_{{\rm d}}]_{ij} \sim \mathcal{C}\mathcal{N}(0,1)$, $[\bm{\tilde{h}^{\textrm{NLOS}}}_{{\rm r},k}]_{ij} \sim \mathcal{C}\mathcal{N}(0,1)$ and $[\bm{\tilde{G}}_{{\rm r},k}^{\textrm{NLOS}}]_{ij} \sim \mathcal{C}\mathcal{N}(0,1)$.


\subsection{Pilot Transmission Protocol}

When there is a localization request, the BS receives a sequence of $T$ uplink pilot symbols from the UE over $T$ time frames. For each pilot frame, the sensing vectors (which include the BS beamforming vector and multiple RISs reflection coefficients) are actively reconfigured to produce a different measurement. 
Given an RIS codebook and a BS beamforming codebook, the sequence of RIS codewords and BS beamforming codewords can either be randomly drawn from the codebook or be strategically selected from the respective codebook to enable better measurements. 
For the $t$-th time frame, let $\bm{\theta}^{(t)}_k$ be the codeword drawn from the codebook $\bm{\Theta}^{\rm RIS}$ for the $k$-th RIS, $\bm{w}^{(t)}$ be the codeword drawn from the codebook $\bm{W}^{\rm BS}$ for the BS, 
and $ x^{(t)} \in \mathbb{C}$ be the pilot symbol to be transmitted from the UE to the BS. 
As shown in Fig.~\ref{fig.setup_intro}, the BS receives a combined signal from the direct link and the RISs-aided link, 
so the received pilot at the BS after RF chain processing is a function of the analog beamforming codeword $\bm{w}^{(t)}$ and the set of $K$ RIS codewords at the $t$-th time frame $[\bm{\theta}^{(t)}_{k}]_{k=1}^{K}$ as follows:
\begin{equation}\label{ris_pilot}
\begin{split}
    y^{(t)}(\bm{w}^{(t)},& 
	[\bm{\theta}^{(t)}_{k}]_{k=1}^{K})=  \\ & 
\sqrt{P_u}(\bm{w}^{(t)})^\top \biggl( \bm{h}_{\rm d} + \sum_{k=1}^K\bm{H}_{{\rm c},k}^\top \bm{\theta}^{(t)}_{k}\biggl) x^{(t)}+ n^{(t)}, 
\end{split}
\end{equation}
where $P_u$ is the uplink transmit power, 
and $n^{(t)} \sim\mathcal{C}\mathcal{N}(0,\sigma_u^2)$ is the uplink additive white Gaussian noise (AWGN) for $t = 0,\cdots, T-1$. Here, $\bm{H}_{{\rm c},k}$ denotes the cascade channel between the BS and the UE through the reflection at the $k$-th RIS:
\begin{equation}
    \bm{H}_{{\rm c},k} = \text{diag}(\bm{h}_{{\rm{r}},k})\bm{G}_{{\rm{r}},k}^\top\in\mathbb{C}^{N\times M}.
\end{equation}

\subsection{Problem Formulation}

Under the active sensing framework, the goal of the localization problem is to 
estimate the unknown user position $\bm{p}$ based on $T$ observations $[{y}^{(t)}]_{t=0}^{T-1}$, by actively drawing RIS codewords and BS beamforming codewords from the two codebooks and reconfiguring the RISs and the BS accordingly for each observation. 
In this work, we propose to design beamforming codebooks and the associated codeword selection mechanism to enable sequential drawing of the codewords to progressively narrow the search area by using more focused beams.
Specifically, we consider the following codebook construction setup. 
Let $\bm{\Psi}$ denote some underlying parameters of the environment, which include critical information such as the channel characteristics (path loss, fading distributions) and the geometry of the service area. 
It is advocated in \cite{env_aware, codebook_learning_RL} that the codebook should be designed in accordance with the site-specific parameters for improved network performance. 
Conceptually, the RIS codebook and BS beamforming codebook construction problem can be thought of as
\begin{equation}\label{function_h}
    \left[ \bm{\Theta}^{\rm RIS}, \bm{W}^{\rm BS} \right]   = \mathcal{H}(\bm{\Psi}),
\end{equation}
where $\mathcal{H}:\mathbb{C}^{{\rm dimension}(\bm{\Psi})} \rightarrow \mathbb{C}^{N\times V} \times \mathbb{C}^{M\times B}$ denotes the mapping from environment parameters to an RIS codebook with $V$ RIS codewords, and a BS beamforming codebook with $B$ beamforming codewords. 
All codewords in the two codebooks satisfy the unit modulus constraint. 
Formulation (\ref{function_h}) stresses that the codebook should be designed as a function of the environmental conditions encapsulated by $\bm{\Psi}$, allowing the codebook to adapt to the site-specific characteristics of the environment.



In the $t$-th time frame, the BS draws a codeword $\bm{w}^{(t+1)}$ from the BS beamforming codebook and draws $K$ codewords $[\bm{\theta}^{(t+1)}_{k}]_{k=1}^{K}$ from the RIS codebook based on the existing observations. These codewords are used to configure the BS and the RISs via the controller 
to make the next measurement ${y}^{(t+1)}$ in the next time frame. We cast the BS beamforming codeword and RIS codewords selection process as a mapping from the historical measurements to the indices of the selected codewords:
\begin{subequations}\label{function_theta}
\begin{align}
       \left[\: j,  i_1, i_2, \cdots, i_K  \right]  & =  \mathcal{G}^{(t)}([{y}^{(\tau)}]_{\tau=0}^{t}) , \label{function_theta1} \\
        [i_k]_{k=1}^K   & \in \{1, \cdots, V\},\\
            j   &\in \{1, \cdots, B\},  
    \label{function_theta2} 
\end{align}
\end{subequations}
where $\mathcal{G}^{(t)}:\mathbb{C}^{t+1} \rightarrow  (\mathbb{N} )^{K+1}  $ denotes the mapping from the received pilot measurements made so far to the index of the BS beamforming codeword and the indices of the $K$ RIS codewords. 
The selected codewords are drawn from the respective codebooks as follows:
\begin{subequations}
\begin{align}  
    \bm{w}^{(t+1)} & =   [\bm{W}^{\rm BS}]{(:,j)},\\
    \bm{\theta}^{(t+1)}_k & =   [\bm{\Theta}^{\rm RIS}]{(:,i_k)}, ~k = 1,\cdots , K.
\end{align}
\end{subequations}
Here, we use the notation $[\bm{W}^{\rm BS}]{(:,j)}$ and $[\bm{\Theta}^{\rm RIS}]{(:,i_k)} $ to denote the $j$-th column of codebook $\bm{W}^{\rm BS}$ and 
$i_k$-th column of codebook $\bm{\Theta}^{\rm RIS}$ respectively. Hence, the $j$-th codeword is drawn for the BS and the $i_k$-th codeword is drawn for the $k$-th RIS. 
For the first pilot, given that no prior observation is available, the first BS beamforming codeword and RIS codewords are produced via function mapping $ \mathcal{G}^{(-1)}(\emptyset)$, which takes an empty set as input and always select the same codewords from the codebooks for the BS and RISs. 

After receiving all observations across $T$ time frames, an estimation of the UE position $\hat{\bm{p}}$ is made based on the past observations:
\begin{equation} \label{function_f}
    \hat{\bm{p}}  = \mathcal{F}([{y}^{(t)}]_{t=0}^{T-1}),
\end{equation}
where $\mathcal{F}:\mathbb{C}^{T} \rightarrow  [{x},{y},{z}]^\top$ denotes the mapping from past received pilots to the estimated UE position.

Given the codebooks, the codeword selection problem for active sensing can be characterized as follows:
\begin{subequations} \label{prob_formulation}
\begin{align}
\underset{\scriptsize \begin{array}{ll}  \{\mathcal{G}^{(t)}(\cdot)\}_{t=0}^{T-1}, \mathcal{F}(\cdot)  \end{array} }{\textrm{minimum}} & \mathbb{E}\left[ \| \hat{\bm{p}}- \bm{p} \|_2^2 ~| ~ \bm{\Theta}^{\rm RIS}, \bm{W}^{\rm BS} \right] \label{prob_formulation_obj} \\ 
\textrm{subject to} 
\;\;\quad  & \quad \quad \quad ( {\rm \ref{function_theta} } ), \;  ({ \rm \ref{function_f}}).
\end{align}
\end{subequations}

Hence, the goal here is to construct codebooks as in (\ref{function_h}), such that once used to perform active sensing in (\ref{prob_formulation}), low localization error can be achieved. The analytical design of such codebooks and the codeword selection mechanism that enables sequential drawing of codewords is difficult. 
To make the problem tractable, some prior works resort to a hierarchical codebook based approach, where the codebook is constructed based on some channel models, and the sequence of codewords is selected based on heuristics that gradually narrow the search range\cite{hierarchical1,hejiguang, hiarchicalwuqq,hierarchical2layer}. 
However, a hierarchical codebook based method still employs exhaustive search within the narrowed search range to identify the optimal codeword. Moreover, 
such a hierarchical codebook based method is greedy in the sense that it only exploits the direction of interest, without exploring alternative directions. An optimal codebook should strike a balance between exploitation and exploration in the beam space. 


In this paper, we employ a neural network to parameterize the function mapping $\mathcal{H}(\cdot)$ in (\ref{function_h}) to construct the BS beamforming codebook and the RIS codebook. The same neural network is also used to adaptively select the sequence of BS beamforming codewords and RIS codewords for sensing as more measurements become available, in effect, by solving problem (\ref{prob_formulation}).

\begin{figure}[t]
\centering
\includegraphics[width=0.7\columnwidth]{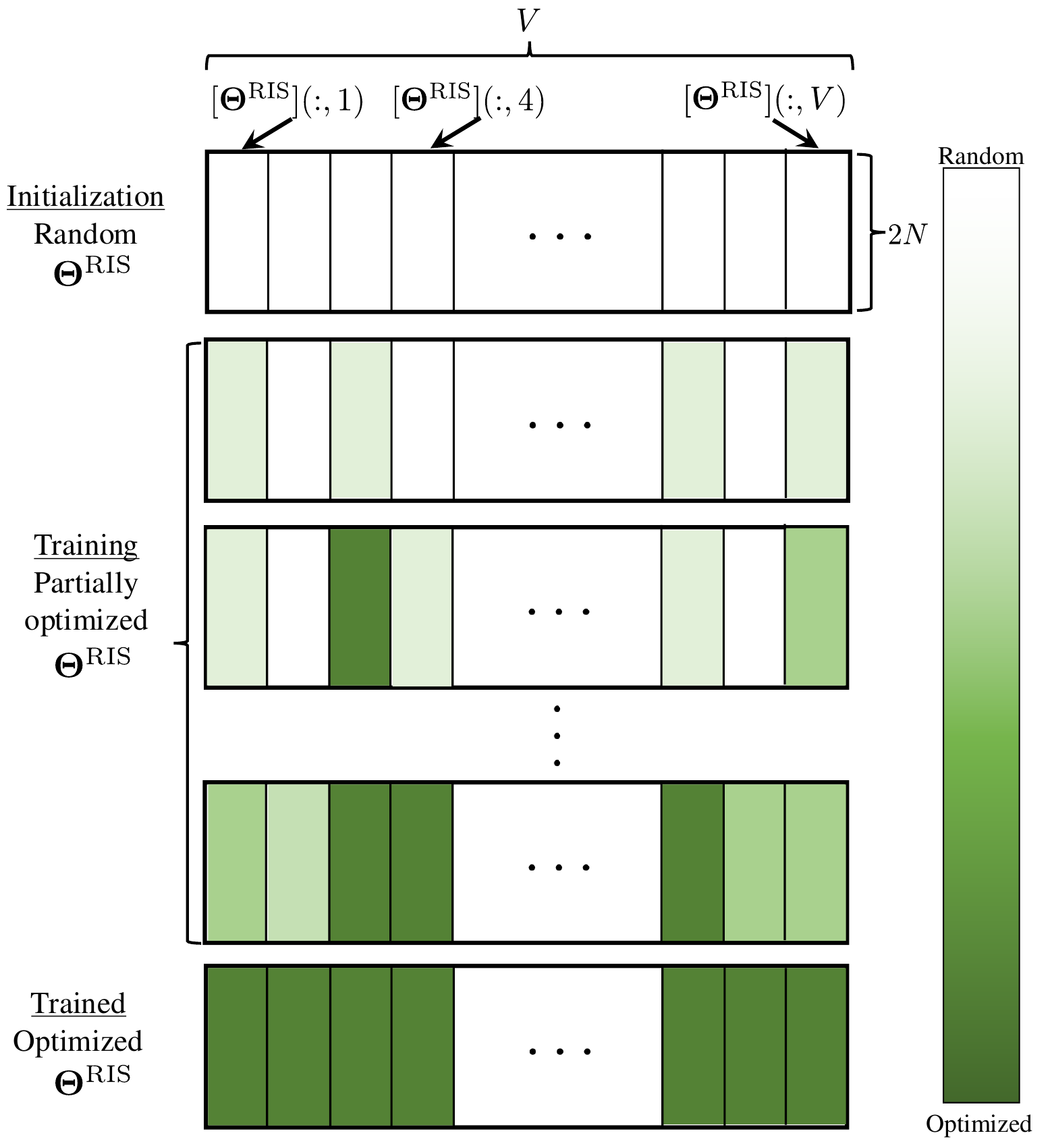}%
\caption{Illustration of the evolution of RIS codebook as training progresses. }%
\label{fig.evolve}%
\end{figure}

\begin{figure*}[!t]
    \centering
  \includegraphics[width=\textwidth]{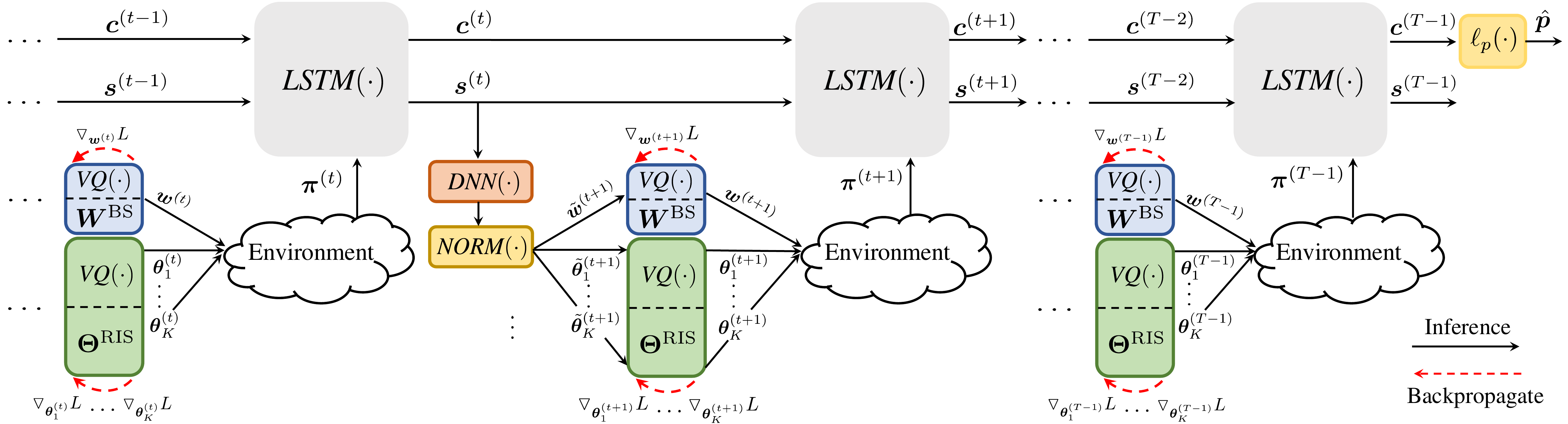}%
  \caption{Proposed codebook learning framework for active sensing.}
  \label{fig.lstmcell}%
\end{figure*}

\section{Proposed Codebook Learning Solution}
\label{sec.rnn}


The conventional codebook-based methods often separately treat the codebook construction and the codeword selection problems. Typically, a codebook is constructed first based on some channel models, 
then based on the fixed codebook, exhaustive search or some heuristics is performed for codeword selection. 
In this paper, we propose a data-driven approach to jointly learn to construct codebooks and to recursively select codewords from the respective codebook based on measurements received, without exhaustive beam training and without assuming a specific channel model. Here, the BS beamforming codebook and the RIS codebook are learned through the codeword selection process where 
for each training sequence, a subset of codewords from the respective codebooks are selected and updated via \emph{forward propagation} and \emph{backward propagation}.  After many training sequences, all codewords in both codebooks are updated repeatedly to reach an equilibrium.

The proposed learning approach is realized by integrating key ideas from VQ-VAE \cite{discretereplearning}, a generative model based on the principle of VQ, into the LSTM network in \cite{twcactive}. 
The use of LSTM network is vital because the network is capable of extracting the temporal dependencies from a sequence of temporal measurements and based on the extracted features, automatically constructing an information vector of fixed dimension. However, the use of LSTM network alone is not sufficient to learn the mapping from an information vector to a codeword index, because  
due to the categorical nature of the codebook, the design space of the RIS reflection coefficients and the BS beamforming vector is discrete. 
Hence, we use the idea of VQ to help design the mapping from measurements made to a codeword index during \emph{forward propagation} by quantizing the LSTM-designed sensing vector to a codeword from the codebook. However, the VQ operation lacks a gradient, which renders  
\emph{backward propagation} ineffective for updating the LSTM network. To address this issue, we use the idea of gradient approximation to estimate the unknown gradient to allow for weight updates. 
Lastly, we design a composite loss function to update the $T$ selected codewords from the BS beamforming codebook and the $TK$ selected codewords from the RIS codebook for each training sequence until all codewords in the two codebooks have been updated, as illustrated in Fig.~\ref{fig.evolve}.

\subsection{Codebook Initialization}
The RIS codebook $\bm{\Theta}^{\rm RIS}$ and the BS beamforming codebook $\bm{W}^{\rm BS}$ consist of $V$ codewords ($V$-way categorical) and $B$ codewords ($B$-way categorical), respectively.
The entries in the codebooks are
randomly initialized according to $\mathcal{C}\mathcal{N}(0,2\pi)$.  
Each codeword undergoes an element-wise normalization as follows to ensure unit modulus constraint
\begin{subequations} \label{unitmodu}
\begin{align}
    [\bm{\theta}]_n = ~  & \dfrac{[\mathcal{R}(\bm{{\theta}})]_{n}}{\sqrt{[\mathcal{R}(\bm{{\theta}})]_{n}^2+[\mathcal{I}(\bm{{\theta}})]_{n}^2}}  +  j\dfrac{[\mathcal{I}(\bm{{\theta}})]_{n}}{\sqrt{[\mathcal{R}(\bm{{\theta}})]_{n}^2+[\mathcal{I}(\bm{{\theta}})]_{n}^2}}, \\
    [\bm{w}]_m = ~ &  \dfrac{[\mathcal{R}(\bm{w})]_{m}}{\sqrt{[\mathcal{R}(\bm{w})]_{m}^2+[\mathcal{I}(\bm{w})]_{m}^2}} 
    +  j\dfrac{[\mathcal{I}(\bm{w})]_{m}}{\sqrt{[\mathcal{R}(\bm{w})]_{m}^2+[\mathcal{I}(\bm{w})]_{m}^2}}. 
\end{align}
\end{subequations}

Since the existing deep learning software packages provide limited support for complex-valued operations, we concatenate the real and imaginary components of the codeword such that the concatenated RIS codebook has $2N\times V$ trainable entries in total, and the concatenated BS beamforming codebook has $2M\times B$ trainable entries in total.

\subsection{Forward Propagation}

The proposed neural network architecture is depicted in Fig.~\ref{fig.lstmcell}. 
The hidden state vector $\bm{s}^{(t-1)}$ and the cell state vector $\bm{c}^{(t-1)}$ of the LSTM network are the information vectors that contain information about the temporal measurement before the $t$-th time frame. 
At the $t$-th time frame, the LSTM cell accepts new features $\bm{\pi}^{(t)}$ as input to update the hidden state vector and the cell state vector:
\begin{equation}\label{lstmupdate}
    (\bm{c}^{(t)}, \bm{s}^{(t)}) = \textit{LSTM}(\bm{\pi}^{(t)},  \bm{c}^{(t-1)}, \bm{s}^{(t-1)} ),
\end{equation}
where $\bm{\pi}^{(t)}$ is the concatenated real and imaginary part of received pilots $[\mathcal{R}({y}^{(t)}), \mathcal{I}({y}^{(t)})]$ and the updating rule of $\textit{LSTM}(\cdot)$ is detailed in \cite{twcactive}. 

It is shown in \cite{twcactive} that
the hidden state vector $\bm{s}^{(t)}$ can be used to map to a BS beamforming vector and a set of RIS configurations in a codebook-free setting, but with the introduction of the codebook, the mapping from the hidden state vector to the codeword index is difficult to learn because the set of codewords in the codebook is discrete. We opt for a two-step approach where the hidden state vector is mapped to a BS beamforming vector and a set of RIS configurations, then the BS beamforming vector and the set of RIS configurations are quantized to codewords. 
The mapping from the hidden state vector to the sensing vectors is as follows:
\begin{equation} \label{norm_dnn}
    \left( {\tilde{\bm w}}^{(t+1)}, [{\tilde{\bm \theta}}_k^{(t+1)}]_{k=1}^K  \right)   = \textit{NORM}(\textit{DNN}(\bm{s}^{(t)})),
\end{equation}
where a deep neural network (DNN), denoted as $\textit{DNN}(\cdot)$, maps the hidden state vector to the right representation of information to design the sensing vector \cite{twcactive}, 
and $\textit{NORM}(\cdot)$ enforces unit modulus constraint for the BS beamforming vector and $K$ RIS configurations as in (\ref{unitmodu}). 
The designed sensing vectors are of dimension 
$ {\tilde{\bm \theta}}_k^{(t+1)} \in  \mathbb{C}^{2N} $
and 
$ {\tilde{\bm w}}^{(t+1)}  \in  \mathbb{C}^{2M} $. 
The quantization of the sensing vector to the codeword from the respective codebook is as follows:
\begin{subequations}\label{vq}
    \begin{align}
        \bm{{\theta}}_k^{(t+1)} & = \textit{VQ}({\tilde{\bm \theta}}_k^{(t+1)} ,  \bm{\Theta}^{\rm RIS} ) =   [\bm{\Theta}^{\rm RIS}]{(:,i_k)}, ~k = 1,\cdots , K, \\
    \bm{{w}}^{(t+1)} & = \textit{VQ}({\tilde{\bm w}}^{(t+1)} ,  \bm{W}^{\rm BS} ) =   [\bm{W}^{\rm BS}]{(:,j)}, 
    \end{align}
\end{subequations}
where 
\begin{subequations}
    \begin{align}
        i_k & = \argmin_{ v \in \{1, \cdots, V \} }  \lVert       {\tilde{\bm \theta}}_k^{(t+1)} -    [\bm{\Theta}^{\rm RIS}]{(:,v)}   \rVert_2 , ~k = 1,\cdots , K, \\
        j &= \argmin_{ b \in \{1, \cdots, B \} }  \lVert       {\tilde{\bm w}}^{(t+1)} -    [\bm{W}^{\rm BS}]{(:,b)}   \rVert_2. 
    \end{align}
\end{subequations}
Hence, the BS uses the $j$-th BS beamforming codeword from $\bm{W}^{\rm BS}$ and the $k$-th RIS uses the $i_k$-th RIS codeword from $\bm{\Theta}^{\rm RIS}$ to 
obtain the pilot measurement for the next time frame. We point out that searching through the codebook for an optimal codeword via VQ can be parallelized and require no additional pilot training, as opposed to existing methods which perform beam-training with every codeword in the codebook to identify the optimal codeword producing the largest SNR.

At each time frame, we select one BS beamforming codeword and $K$ RIS codewords to obtain a measurement. After collecting all measurements in $T$ time frames, 
we obtain the estimated UE position $\hat{\bm{p}}$
based on the final cell state $\bm{c}^{(T-1)}$ via a fully connected neural network $\ell_p(\cdot)$ as follows:
\begin{equation}\label{ellp}
    \hat{\bm{p}} = \ell_p (\bm{c}^{(T-1)}).
\end{equation}

\subsection{Backward Propagation}

The use of VQ in forward propagation is simple and effective, yet it leaves complications for backward propagation as there is no gradient defined for operation (\ref{vq}). During backpropagation, the gradients of the loss function with respect to the weights prior to the VQ operation cannot be determined. Consequently, 
the weights of $\textit{DNN}(\cdot)$ in (\ref{norm_dnn}) cannot be updated meaningfully to reduce the loss and thereby unable to design better 
pre-quantized sensing vectors $[{\tilde{\bm \theta}}_k^{(t+1)}]_{k=1}^K$ and ${\tilde{\bm w}}^{(t+1)}$.

To train those weights, 
since the pre-quantized and quantized sensing vectors share the same dimensionality ($2N$ for RIS reflection coefficients, $2M$ for BS beamforming vector), 
we approximate the gradient by copying the known gradients of the quantized sensing vectors, ${{\bm \theta}}_k^{(t+1)}$ and ${{\bm w}}^{(t+1)}$, to be the gradient of the pre-quantized sensing vectors, ${\tilde{\bm \theta}}_k^{(t+1)}$ and ${\tilde{\bm w}}^{(t+1)}$: 
\begin{subequations}\label{gradientapprox}
    \begin{align}
    \triangledown_{ {\tilde{\bm \theta}}_k^{(t+1)}} L & = \triangledown_{ {{\bm \theta}}_k^{(t+1)}} L , ~k = 1,\cdots , K,\\
    \triangledown_{ {\tilde{\bm w}}^{(t+1)}} L & = \triangledown_{ {{\bm w}}^{(t+1)}} L .
\end{align}
\end{subequations}
Here, the gradients $\triangledown_{ {\tilde{\bm \theta}}_k^{(t+1)}} L $ and $\triangledown_{ {\tilde{\bm w}}^{(t+1)}} L $ are approximations, but they still contain useful information for the $\textit{DNN}(\cdot)$ in (\ref{norm_dnn}) for updating its weights to design better RIS configurations and BS beamforming vector based on the hidden state vector $\bm{s}^{(t)}$.  




The realization of the gradient approximation in (\ref{gradientapprox}) can be achieved by redefining the chosen codewords $[{\bm \theta}_k^{(t+1)}]_{k=1}^{K}$ and ${\bm w}^{(t+1)}$ after the VQ operation in (\ref{vq}) as follows:
\begin{subequations}\label{redefine}
    \begin{align}
         {{\bm \theta}}_k^{(t+1)} & =   {\tilde{\bm \theta}}_k^{(t+1)} -   \textit{SG}(  {{\bm \theta}}_k^{(t+1)}  -{\tilde{\bm \theta}}_k^{(t+1)} ), ~k = 1,\cdots , K, \\
          {{\bm w}}^{(t+1)} & =   {\tilde{\bm w}}^{(t+1)} -   \textit{SG}(  {{\bm w}}^{(t+1)}  -{\tilde{\bm w}}^{(t+1)} ).
    \end{align}
\end{subequations}
Here, as proposed in \cite{discretereplearning}, we use stop-gradient operator $\textit{SG}(\cdot)$ which acts as an identity operator in forward propagation and has zero partial derivatives. As a result, the operand of the $\textit{SG}(\cdot)$ operator is disregarded for computing gradients during backpropagation, which effectively achieves (\ref{gradientapprox}).


\begin{algorithm}[!t]
    \caption{VQ-C Algorithm} \label{algo1}
    \SetAlgoLined
    \SetKwInOut{Input}{input}
    \SetKwInOut{Output}{output}
    \DontPrintSemicolon
    \Input{Initialize RIS codebook 
    $\bm{\Theta}^{\rm RIS}$, \linebreak
    Initialize BS beamforming codebook 
    $\bm{W}^{\rm BS}$, \linebreak
    $Q$ training sequences.
    .}
    
    \Output{Optimized RIS codebook $\bm{\Theta}^{\rm RIS}$, \linebreak
    Optimized BS beamforming codebook $\bm{W}^{\rm BS}$, \linebreak
    Optimized LSTM and DNN network.}

    \For { $q = 1,2, \cdots, Q$ } {

    \hskip-0.5em \emph{Forward propagation}:
    
    \For  {$t = 0,1, \cdots, T-1$} {

        UE transmits pilot $x^{(t)}$;
        
        BS receives pilot $y^{(t)}$ through the RISs;

        Update state vectors $\bm{c}^{(t)}$ and $\bm{s}^{(t)}$ as in (\ref{lstmupdate});

        Design sensing vectors ${\tilde{\bm w}}^{(t+1)}, {\tilde{\bm \theta}}_k^{(t+1)}$ as in (\ref{norm_dnn});

        Quantize sensing vectors to codewords as in (\ref{vq}); \label{algo:vq}

        Redefine selected codewords as in (\ref{redefine});
        
        

    }
    Estimate position $\hat{\bm{p}}$ as in (\ref{ellp});
    
    Compute MSE loss with ${\bm{p}}$ as in (\ref{lossfn});
    
    Find codebook loss of chosen codewords in step \ref{algo:vq}.
    
    \hskip-0.5em \emph{Backward propagation}:

    Update $\textit{LSTM}(\cdot)$ and $\textit{DNN}(\cdot)$ via MSE loss;
        
    Update $T$ selected BS beamforming codewords via BS beamforming codebook loss.
            
    Update $TK$ selected RIS codewords via RIS codebook loss.
    
    }
    Return  $\bm{\Theta}^{\rm RIS}$,  $\bm{W}^{\rm BS}$, $\textit{LSTM}(\cdot)$, $\textit{DNN}(\cdot)$.
    
    \label{alg:PoEG}
\end{algorithm}

\subsection{Loss Function}

We use a composite loss function that contains three terms to train the two codebooks and the LSTM network. The three terms are the mean-squared error (MSE) loss, BS beamforming codebook loss, and RIS codebook loss respectively: 
\begin{equation} \label{lossfn}
    L = \mathbb{E}\left[ \| \hat{\bm{p}}- \bm{p} \|_2^2 \right] + \sum_{t=0}^{T-1} \alpha^{(t+1)} + \sum_{k=1}^{K} \sum_{t=0}^{T-1} \beta^{(t+1)}_k.
\end{equation}

To train the LSTM network, MSE loss is used to minimize the average MSE between the estimated position $\hat{\bm{p}}^{(T)}$ and the true position $\bm{p}$ by training the weights of the $\textit{LSTM}(\cdot)$, $\textit{DNN}(\cdot)$ and $\ell_p(\cdot)$ functions through the gradients estimator in (\ref{gradientapprox}). 

The middle term of (\ref{lossfn}) is introduced to train the BS beamforming codewords and it is defined as follows:
\begin{equation}
    \alpha^{(t+1)} =   \| {\textit{SG}}(          {\tilde{\bm w}}^{(t+1)}   )  - {{\bm w}}^{(t+1)}  \|_2^2  
      + \| {\tilde{\bm w}}^{(t+1)}  - {\textit{SG}}( {{\bm w}}^{(t+1)}  ) \|_2^2.
\end{equation}
Here, $\alpha^{(t+1)}$ includes the codeword loss and the commitment loss. Specifically, given that the operand of $\textit{SG}(\cdot)$ is fixed and not being updated, the codeword loss aims to move the selected BS beamforming codeword (i.e., $\bm{w}^{(t+1)}$ as in (\ref{vq})) toward a configuration with growing similarity with  
the ${\tilde{\bm w}}^{(t+1)}$ component of $\textit{DNN}(\cdot)$ output. 
To ensure that the ${\tilde{\bm w}}^{(t+1)}$ component of $\textit{DNN}(\cdot)$ output commits to a codeword, the commitment loss aims to move the ${\tilde{\bm w}}^{(t+1)}$ component of $\textit{DNN}(\cdot)$ output toward the selected codeword i.e., ${{\bm w}}^{(t+1)}$. 
The joint effect of the two terms of $\alpha^{(t+1)}$
moves the pre-quantized BS beamforming vector (${\tilde{\bm w}}^{(t+1)}$ component of $\textit{DNN}(\cdot)$ output) and the selected codeword ${{\bm w}}^{(t+1)}$ closer in $\ell_2$ distance. 

Similarly, the third term of (\ref{lossfn}) is introduced to train the RIS codewords:
\begin{equation}
    \beta^{(t+1)}_k =  \| {\textit{SG}}(          {\tilde{\bm \theta}}_k^{(t+1)}   )  - {{\bm \theta}}_k^{(t+1)}  \|_2^2  
      + \| {\tilde{\bm \theta}}_k^{(t+1)}  - {\textit{SG}}( {{\bm \theta}}_k^{(t+1)}  ) \|_2^2 .
\end{equation}
Here, $\beta^{(t+1)}_k$ also includes the codeword loss and the commitment loss respectively to move the pre-quantized RIS patterns (${\tilde{\bm \theta}}_k^{(t+1)}$ component of $\textit{DNN}(\cdot)$ output) and the selected RIS codeword ${{\bm \theta}}_k^{(t+1)}$ closer in $\ell_2$ distance.

The loss function (\ref{lossfn}) enables the update of $T$ codewords from the BS beamforming codebook and $TK$ codewords from the RIS codebook per training sequence. 
After many training sequences, the two codebooks are finalized when all codewords in the codebooks have been updated to reach an equilibrium. The overall algorithm is summarized in Algorithm \ref{algo1}.

\begin{figure}[t]
    \centering
\includegraphics[width=\columnwidth]{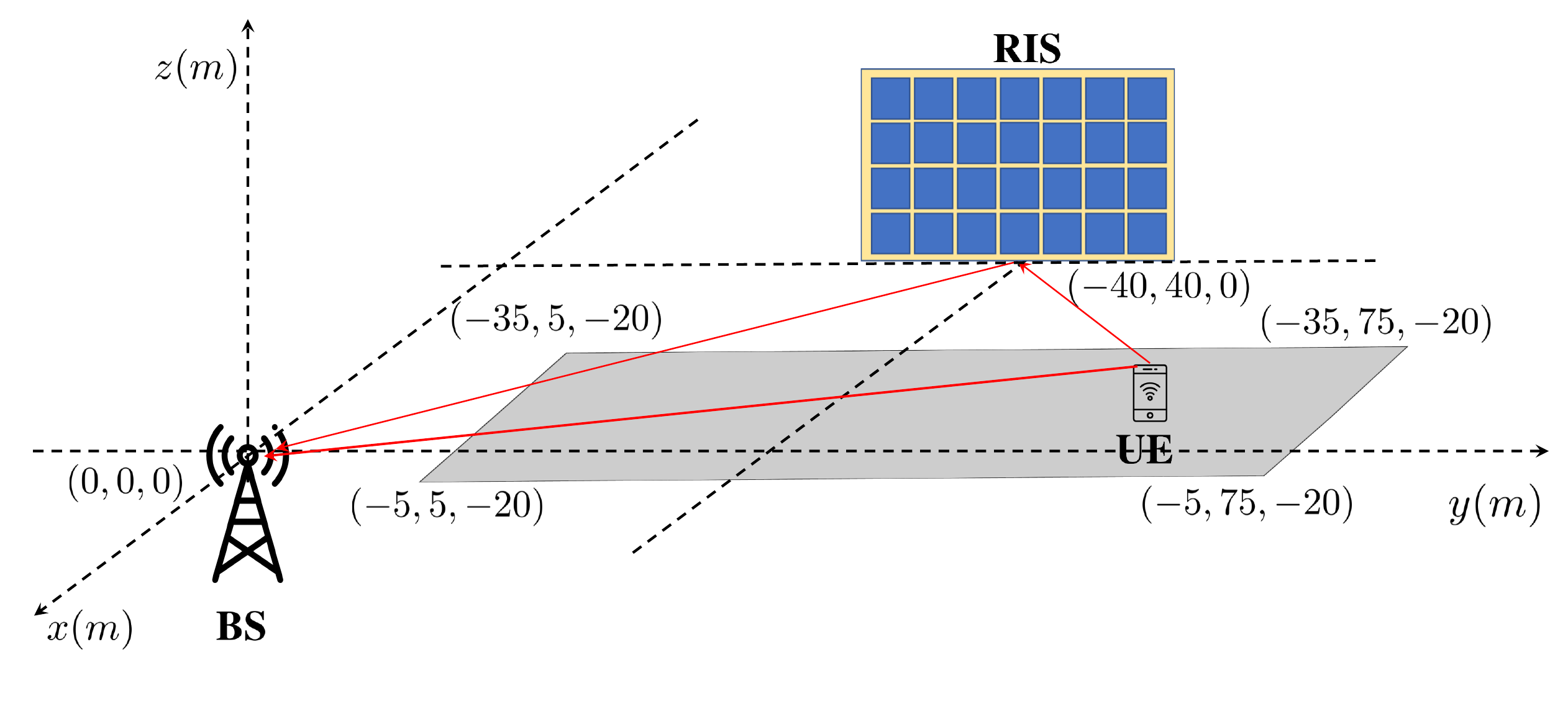}%
\caption{Simulation environment for localization in a single-RIS SISO network.} 
\label{fig.SISOsetup_simulation}%
\end{figure}

\section{Numerical Results for Single-RIS SISO Network}
\label{sec.numerical_irs_single}
We begin with a setting where only $K=1$ RIS is deployed in a SISO network to aid localization, as shown in Fig.~\ref{fig.SISOsetup_simulation}. Here, the BS is equipped with only a single antenna, so there is no BS beamforming codebook, only the RIS codebook. The case with multi-RIS MISO network is considered in Section~\ref{sec.numerical_irs_multi}.
We consider a system setup where the BS is placed at $\bm{p}^{\rm BS} = (0m,0m,0m)$ and the $8\times8$ RIS is placed at $\bm{p}^{\rm RIS} = (-40m,40m,0m)$. 
The UE locations $\bm{p}$ are generated uniformly within a rectangular area on the $x$-$y$ plane $(20m\pm15m, 0m\pm35m, -20m)$.
This is the same simulation setting as in \cite{twcactive}.

\subsection{Channel Model} \label{siso_singleris_environment}

Recall that $\bm{h}_{\rm d}$, $\bm{G}_{{\rm r},k} $ and $\bm{h}_{{\rm r},k} $ denote the direct channel between BS/UE, and the reflection channel between BS/$k$-th RIS, and $k$-th RIS/UE respectively. The channels follow a Rician fading channel model. The path loss component of the channel contains distance information, and the LOS component of the channel contains angular information about the location of the UE.



Specifically, $\bm{\tilde{h}}_{{\rm r},k}^{\rm LOS}$ is a function of the $k$-th RIS's location and the UE location. We use $\mu^{\rm RIS}_k$ and $\gamma^{\rm RIS}_k$ to denote the azimuth and elevation angles-of-arrival (AoA) from the UE to the $k$-th RIS. We can express the LOS component of $\bm{{h}}_{{\rm r},k}$ as
\begin{equation}\label{ris_to_ue}
    \bm{\tilde h}_{{\rm r},k}^{\rm LOS} = \bm{u}^{\rm RIS}(\mu_k^{\rm RIS}, \gamma_k^{\rm RIS}),
\end{equation}
\noindent where the steering vector associated with the $n$-th element of the RIS $k$ is given by \cite{taojournal} \cite{multiAIRS}
\begin{align*}
\hspace{-0.6em} [   \bm{u}^{\rm RIS } &(\phi_k^{\rm RIS}, \psi_k^{\rm RIS})]_n = \\
&
e^{\scriptsize j \frac{2\pi d_{\rm R}}{\lambda_c} \{ v_1(n, C_k) {\rm sin}(\mu_k^{\rm RIS}){\rm cos}(\gamma_k^{\rm RIS}) + v_2(n, C_k) {\rm sin}(\gamma_k^{\rm RIS}) \}}.\label{eqn_1} \numberthis 
\end{align*}
%
Here, we use $d_{\rm R}$ to denote the distance between two reflective elements of the RIS, $\lambda_c$ to denote the carrier wavelength, and $v_1(n,C) = {\rm mod}(n-1,C)$ and $v_2(n, C) = \lfloor{\frac{n-1}{C}}\rfloor$. We use $C$ to denote the number of columns of the RISs. 
As shown in Fig.~\ref{fig.setup_intro} and Fig.~\ref{fig.SISOsetup_simulation}, we establish the mapping from the set of angles $\{\mu_k^{\rm RIS}, \gamma_k^{\rm RIS} \}$ to the position of UE relative to the $k$-th RIS's position $(x_k^{\rm RIS}, y_k^{\rm RIS}, z_k^{\rm RIS})$ as follows:
\begin{subequations}
\begin{align}
\sin(\gamma_k^{\rm RIS})\cos(\mu_k^{\rm RIS})=\;& \dfrac{x - x_k^{\rm RIS}}{r_k^{\rm UR}},\\
\sin(\gamma_k^{\rm RIS})\sin(\mu_k^{\rm RIS})=\;& \dfrac{y - y_k^{\rm RIS}}{r_k^{\rm UR}},\\
\cos(\gamma_k^{\rm RIS})=\;& \dfrac{z_k^{\rm RIS} - z}{r_k^{\rm UR}},
\end{align}
\end{subequations}
where $r_k^{\rm UR}$ denotes the range between the RIS $k$ and the UE.


We use $\varphi_k^{\rm RIS}$ and $\upsilon_k^{\rm RIS}$ to denote the azimuth and elevation angles-of-departure (AoD) from the $k$-th RIS to the BS, and denote $\gamma_k^{\rm BS}$ as the elevation AoA from the $k$-th RIS to the BS. Hence, 
the LOS component of $\bm{G}_{{\rm r},k}$ is given by 
\begin{equation}
    \bm{\tilde G}_{{\rm r},k}^{\rm LOS} = ~ \bm{u}^{\rm BS}( \gamma_k^{\rm BS}) \bm{u}^{\rm RIS}(\varphi_k^{\rm RIS}, \upsilon_k^{\rm RIS})^{\rm H},
\end{equation}
where the BS steering vector can be written as
\begin{align*}
\hspace{-0.6em} \bm{u}^{\rm BS} & ( \gamma_k^{\rm BS}) = \\  
 & \biggl[1 ,  e^{\scriptsize \begin{array}{ll} j\dfrac{2\pi d_{\rm A}}{\lambda_c} \cos(\gamma_k^{\rm BS}) \end{array}}, \cdots, e^{\scriptsize \begin{array}{ll} j\dfrac{2\pi d_{\rm A}(M-1)}{\lambda_c} \cos(\gamma_k^{\rm BS}) \end{array}}  \biggl]^\top,
 \label{eqn_2} \numberthis 
\end{align*}
where $d_{\rm A}$ denotes the distance between two adjacent BS antenna. 
Finally, we use $ \gamma^{\rm UE}$ to denote the elevation AoA from the UE to the BS, so the LOS component of the direct channel $\bm{h}_{{\rm d}}$ can be written as
\begin{equation}
\bm{\tilde{h}}_{{\rm d}}^{\rm LOS} = \bm{u}^{\rm BS} ( \gamma^{\rm UE}).
\end{equation}
 Similarly, the following mapping based on the set of angles $\{\varphi_k^{\rm RIS},\upsilon_k^{\rm RIS},\gamma_k^{\rm BS} \}$ relative to the BS's position $(x_k^{\rm BS}, y_k^{\rm BS}, z_k^{\rm BS})$ can be established: 
\begin{subequations}
\begin{align}
\cos(\varphi_k^{\rm RIS})\sin(\upsilon_k^{\rm RIS})=\;& \dfrac{x_k^{\rm RIS} - x^{\rm BS}}{r_k^{\rm RB}},\\
\sin(\varphi_k^{\rm RIS})\sin(\upsilon_k^{\rm RIS})=\;& \dfrac{y_k^{\rm RIS} - y^{\rm BS}}{r_k^{\rm RB}},\\
\sin(\gamma_k^{\rm BS})=\;& \dfrac{z_k^{\rm RIS} - z^{\rm BS}}{r_k^{\rm RB}},
\end{align}
\end{subequations}
where $r_k^{\rm RB}$ denotes the range between the BS and the RIS $k$.

In subsequent simulations, we set the Rician factor $\epsilon$ to $10$, and set $ {2\pi d_{A}}/{\lambda_c} = {2\pi d_{R}}/{\lambda_c}$ to $1$ without loss of generality. The path loss models (in dB) of the direct and reflected channel are $32.6+36.7\log(d_1)$ and $30+22\log(d_2)$, respectively, where $d_1$ denotes the distance in the direct path (between UE and BS), and $d_2$ denotes the distance in the reflected path (between BS/$k$-th RIS, between UE/$k$-th RIS)\cite{pathloss}. 
We use $10$ MHz bandwidth with a noise power spectral density of $-170$ dBm/Hz.

\begin{table}[t]
\renewcommand{\arraystretch}{1.7} 
\begin{center}
\captionof{table}{{\sc Parameters of the VQ-C Network.}\label{Tab:parameter}}
\begin{tabular}{| c|c |} 
\hline
\textbf{Label}& \textbf{Dimension}\\
\hline
\shortstack{\:\\${\bm{{r}}}_o,{\bm{{r}}}_i,{\bm{{r}}}_f,{\bm{{r}}}_c$ \\ ${\bm{{u}}}_o,{\bm{{u}}}_i,{\bm{{u}}}_f,{\bm{{u}}}_c$ } &\shortstack{ $512$ \\ \quad \quad }\\

\hline
\quad\shortstack{\: \\$\bm{A}_{1}$}$ \quad\quad\;\;\, $\vline $ \quad\quad \{\bm{A}_l\}_{l=2}^{L}   $   & \quad $512 \times 1024 \quad\quad $\vline $ \; \;\;\quad  1024 \times 1024$ \\
\hline
\quad\shortstack{\: \\ $\bm{\dot{A}}_{L+1}$} $\quad $\vline $ \quad\quad\quad $\shortstack{\: \\ $\bm{\ddot{A}}_{L+1}$}   & \quad $1024 \times 2NK \quad  $\vline $ \, \quad \quad 1024 \times 2 M$ \\
\hline
\shortstack{\: \\ $\{\bm{b}_l\}_{l=1}^{L}  $}   
\vline  \; \shortstack{ \: \\ $\bm{\dot b}_{L+1}$}  \; 
\vline  \;\;\;\;
\shortstack{\: \\ $\bm{\ddot b}_{L+1}$}   & $ 1024 \quad\quad \; $\vline $\quad \; \; 2NK  \;\;\quad $\vline $ \;\quad\quad   2 M$ \\
\hline
$\ell_{p}$   & $ 512 \times 200 \times 200 \times 200 \times 3$\\
\hline
\end{tabular}
\end{center}
\end{table}

\subsection{Proposed vs. Baseline Schemes}

The proposed VQ-C network with a learnable codebook is implemented using parameters from Table \ref{Tab:parameter}. Those parameters originate from \cite{twcactive} and they specify the dimension of $\textit{LSTM}(\cdot)$, $\textit{DNN}(\cdot)$ and $\ell_p (\cdot)$ in (\ref{lstmupdate}), (\ref{norm_dnn}) and (\ref{ellp}) respectively. 
The algorithm requires $T\log_2 V$ signalling bits and $T$ pilots across $T$ time frames to perform localization. We build the model with Tensorflow \cite{tensorflow} and train the model with Adam optimizer \cite{adam}, training it using 2,048,000 training sequences over 2000 epochs. We evaluate its localization performance against the following benchmarks.

\emph{Codebook-free LSTM network \cite{active2023icc, twcactive}}:  The sequence of $T$ uplink RIS configurations is adaptive based on existing measurements in a codebook-free fashion. This scheme needs $T$ pilots and $TN \log_2 B$ signalling bits to configure $N$ RIS elements, each of $B$ possible phase shift values, over $T$ time frames. This is a suitable scheme when the BS-RIS control link is not rate-limited. 

\emph{DNN with random or learned RIS reflection coefficients}:
The DNN designs a sequence of $T$ uplink RIS reflection coefficients that are non-adaptive. 
Here, we can have two different schemes: \romannum{1}) the sequence of $T$ RIS reflection coefficients are randomly generated, or \romannum{2}) the sequence of $T$ RIS reflection coefficients are learned from training data, but is not adaptive based on the measurements made. Scheme (\romannum{2}) is realized in Tensorflow by making the sequence of RIS reflection coefficients trainable during the neural network training phase, allowing them to be tuned according to the channel statistics to minimize the loss function. 
Under both schemes, $T$ pilots are needed to configure the RIS. We use a fully connected DNN to map the received pilots over $T$ time frames $[ \mathcal{R}({y}^{(t)}),\mathcal{I}({y}^{(t)})]_{t=0}^{T-1}$ to an estimated UE position. The dimension of the DNN is $[ 200,200,200,3 ]$.

\emph{Optimizing BCRLB using gradient descent (GD)}\cite{twcactive}:
We design an active sensing strategy by minimizing the Bayesian Cram\'er-Rao lower bound (BCRLB) in each time frame. This involves updating the posterior distribution of the unknown UE position based on existing pilots and accordingly updating the conditional BCRLB in each time frame. We then identify the RIS codeword by optimizing the RIS reflection coefficients with the objective of minimizing the conditional BCRLB, and quantizing the design RIS configuration to a codeword. The codebook consists of a selection of optimized RIS configurations that minimize BCRLB in the codebook-free setting\cite{twcactive}. Here, the posterior distribution of the $30m\times70m$ service area on the $x$-$y$ plane is discretized to $300 \times 700$ blocks to compute the posterior distribution.


\begin{figure}
\centering
\includegraphics[width=\columnwidth]{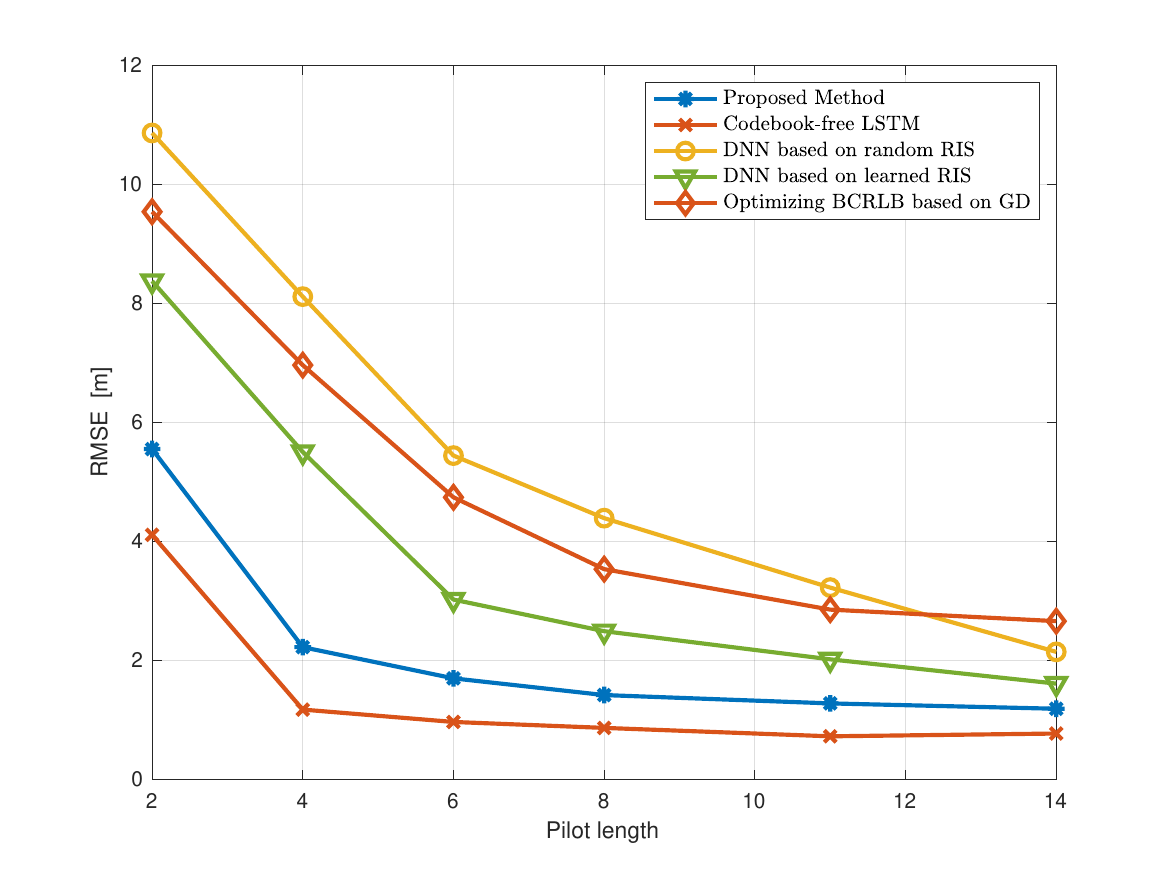}%
\caption{Localization error vs. pilot length with $N=64$, raw SNR  $=25$dB, codebook size $ V =10000$ in a single-RIS SISO network.}%
\label{fig.loc_riscodebook_mse_vs_t}%
\end{figure}

\subsection{Simulation Results}

We examine the localization performance in terms of root-mean-squared error (RMSE), i.e., $\| \hat{\bm{p}}- \bm{p} \|_2$,  with varying numbers of time frames for fixed raw SNR, i.e., $P_u = 10^{{\rm SNR}/10}$. Fig.~\ref{fig.loc_riscodebook_mse_vs_t} shows that the proposed VQ-C approaches the performance of its codebook-free counterpart with $10000$ codewords, which amounts to $14$ bits per control signal to specify the codeword index as opposed to hundreds of bits to express the entire RIS configuration. However, the performance gap is not zero, due to the neural network's inherent limitation in surveying a discrete function space as compared to continuous ones. 
The proposed method consistently outperforms fixed sensing benchmarks with non-adaptive RIS design across various pilot lengths. This implies that the proposed VQ-C network utilizes existing measurements effectively to select a suitable RIS codeword from the codebook for the next time frame to reduce localization error. 
We also point out that the adaptive RIS design based on BCRLB-minimization does not serve as an optimal design to reduce location MSE due to quantization errors and also due to the fact that
the BCRLB can serve as a loose lower bound for the MSE, particularly when the number of observations is few or the SNR is low \cite{crlbbound2}. 

\begin{figure}
\centering
\includegraphics[width=\columnwidth]{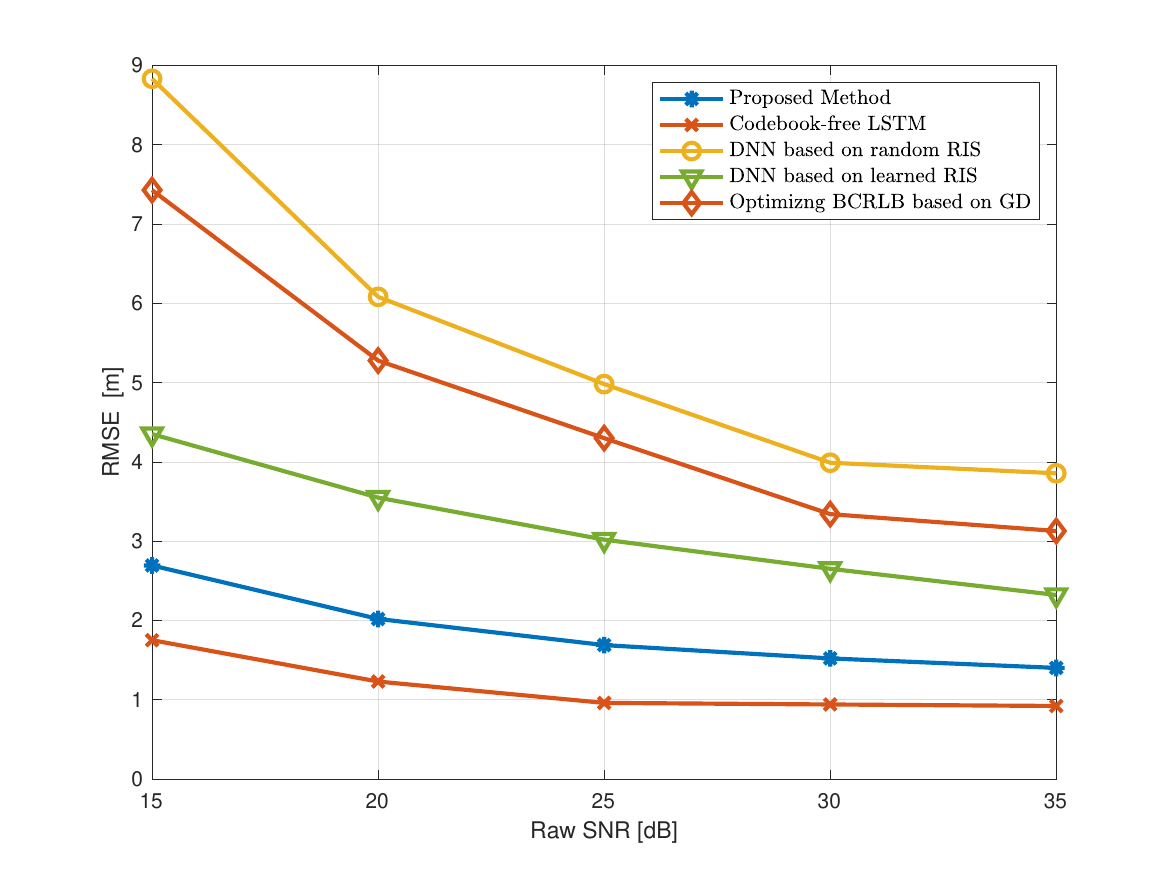}%
\caption{Localization error vs. raw SNR with $T=6$,  codebook size $ V =10000$ in a single-RIS SISO network.}%
\label{fig.codebooksnr}%
\end{figure}

We next study the localization performance of the proposed VQ-C method with varying raw SNR.
From Fig.~\ref{fig.codebooksnr}, we report that the proposed VQ-C algorithm can work with a range of raw SNRs. As the SNR increases, the codebook-based approach is able to match its codebook-free counterpart with some loss in performance, due to discretization of the RIS design space. 

\begin{figure}
\centering
\includegraphics[width=\columnwidth]{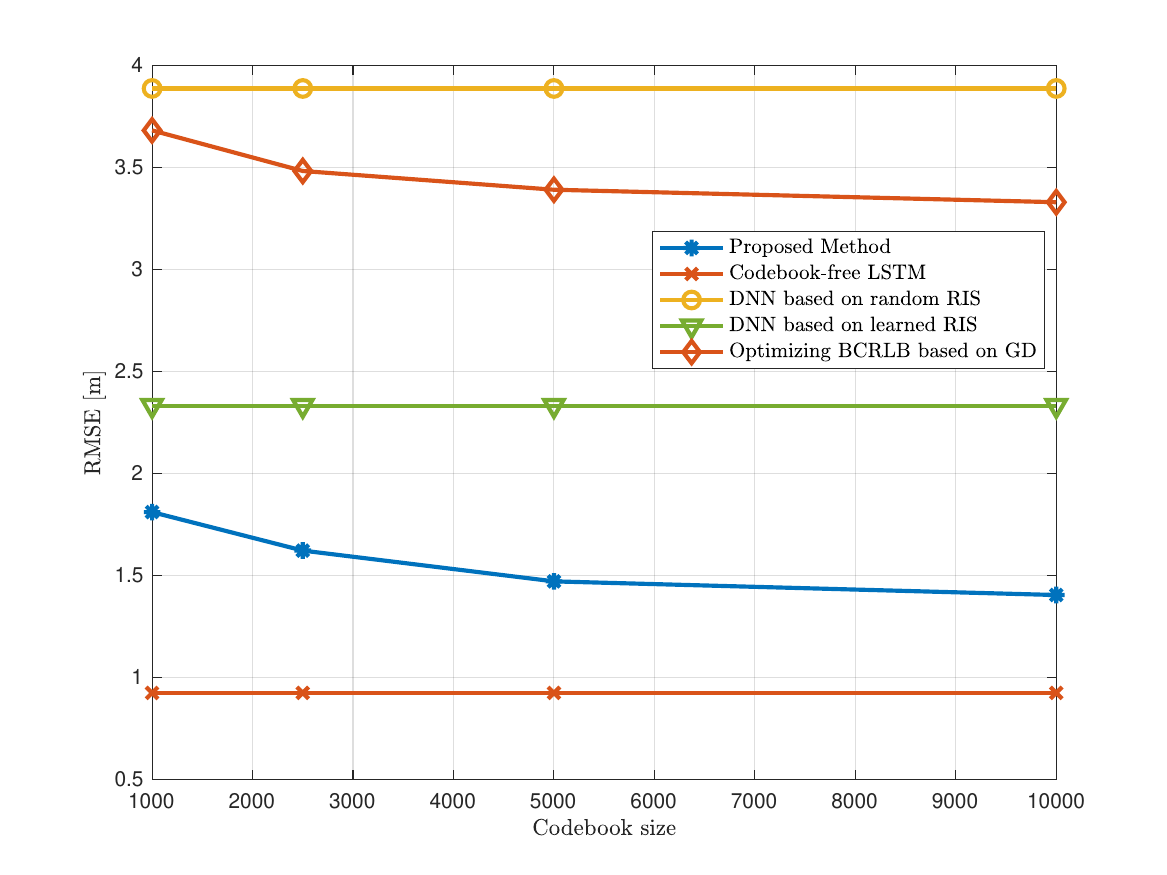}%
\caption{Localization error vs. codebook size with $N=64$, raw SNR  $=35$, $T=6$ in a single-RIS SISO network.}%
\label{fig.codebooksize}%
\end{figure}

We next examine the localization performance of the proposed VQ-C algorithm with varying codebook sizes. From Fig.~\ref{fig.codebooksize}, we observe that the performance of the proposed method approaches the performance of its codebook-free counterpart as the number of codewords in the codebook increases. This is expected as a richer codebook implies a stronger capability to generate a diverse set of reflection channels to aid sensing. 

\begin{figure}[t]
  \centering
  \begin{subfigure}{.49\columnwidth}
    \centering
    \includegraphics[width=\linewidth]{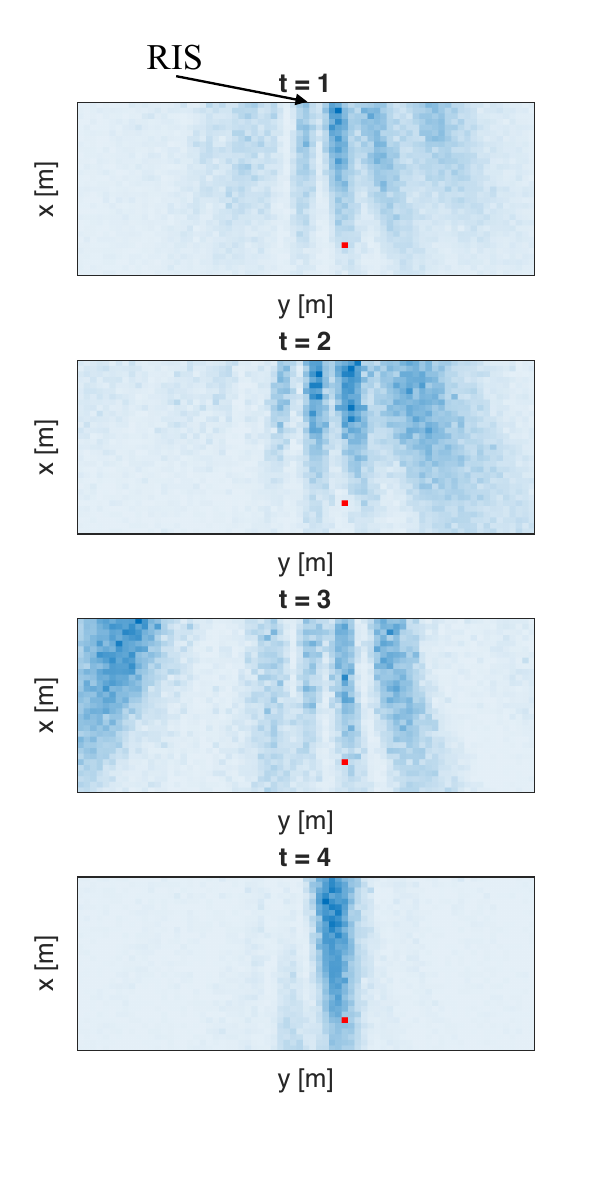}
     \vspace{-3em}
\caption{Pre-quantization.}
    \label{interp_prequan}
  \end{subfigure}%
  \hfill
  \begin{subfigure}{.49\columnwidth}
    \centering
    \includegraphics[width=\linewidth]{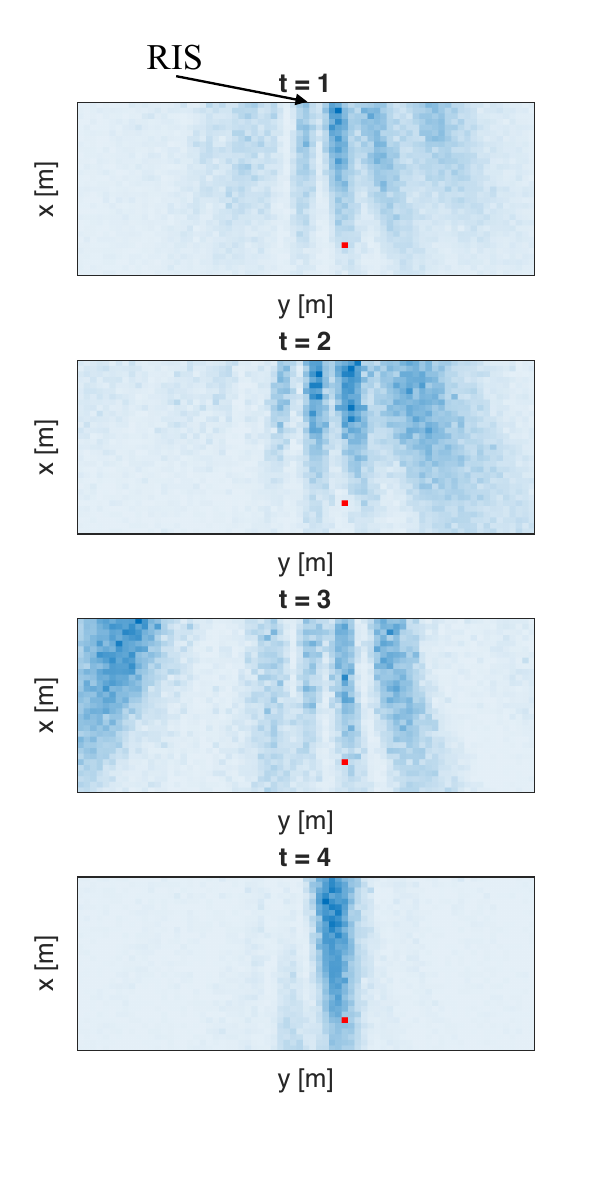}
     \vspace{-3em}
\caption{Post-quantization.}%
    \label{interp_postquan}
  \end{subfigure}%
  \hfill
\caption{RSS distribution from time frame $1$ to $4$ with raw SNR = $25$dB, $T = 4$ in a single-RIS SISO network.}
    \label{loc_ris_interpretation}
\end{figure}

\subsection{Interpretation of the Solution Learned by VQ-C}

We interpret the RIS reflection codebook learned from the proposed VQ-C network. Specifically, we test the trained neural network for a single user with an arbitrary position and use the received signal strength (RSS) distribution (or radio map) as a way to demonstrate qualitatively the correctness of the learned reflection codewords. 
This is achieved by capturing the unquantized RIS reflection coefficients designed by $\textit{LSTM}(\cdot)$ and the corresponding RIS codeword selected from the codebook at each time frame and plotting the RSS value obtained at each $1m \times 1m$ block within the service area across the $x-y$ plane, as depicted in Fig.~\ref{loc_ris_interpretation}. 
We use a red dot to indicate the true position of the UE. 
By comparing Fig.~\ref{loc_ris_interpretation}(\subref{interp_prequan})-(\subref{interp_postquan}), a minimal difference can be perceived from the two sets of radio maps. This implies that the loss function in (\ref{lossfn}) is successful in moving the unquantized and quantized RIS patterns closer in terms of $l_2$ distance. 
We also observe that the RIS adopts a sequence of codewords from the RIS codebook that initially probe the environment with wider beams during the first few time frames to search for the UE, then gradually focus toward the UE with narrower beams. This suggests that the proposed VQ-C network is effective in selecting a meaningful sequence of RIS codewords based on the measurements already made.

\subsection{Complexity Analysis}


Let $T$ denote the number of time frames, $M$ the number of BS antennas, $N$ the number of elements per RIS, and $K$ the total number of RISs. The complexity of the proposed learning-based method involves $T$ neural network inference stages, scaling as ${O}\left[ T \left( D_iZ_i + \Omega + \zeta(KN + M) + \Lambda_{\Theta}K + \Lambda_{W} + D_oZ_o \right) \right]$. Here, $Z_i$ and $Z_o$ represent the dimensions of the input and output layers, respectively, while $\Omega$ accounts for the complexity of the LSTM inference operation during each sensing stage. The term $\zeta$ captures the computational cost of the $(L+1)$-layer neural network used to design sensing vectors, and $\Lambda_{\Theta}$ and $\Lambda_{W}$ represent the complexities of the nearest neighbour search within the RIS and BS beamforming codebooks, respectively. Additionally, $D_i$ and $D_o$ denote the dimensions of the input features, such as pilots, and the output features, respectively. It is worth noting that the nearest neighbour search is parallelizable, and the training and inference processes are also highly parallelizable using modern graphic process units (GPUs), enabling the proposed approach to be implemented efficiently in practice.

\begin{figure}[t]
    \centering
    \vspace{2em}
\includegraphics[width=\columnwidth]{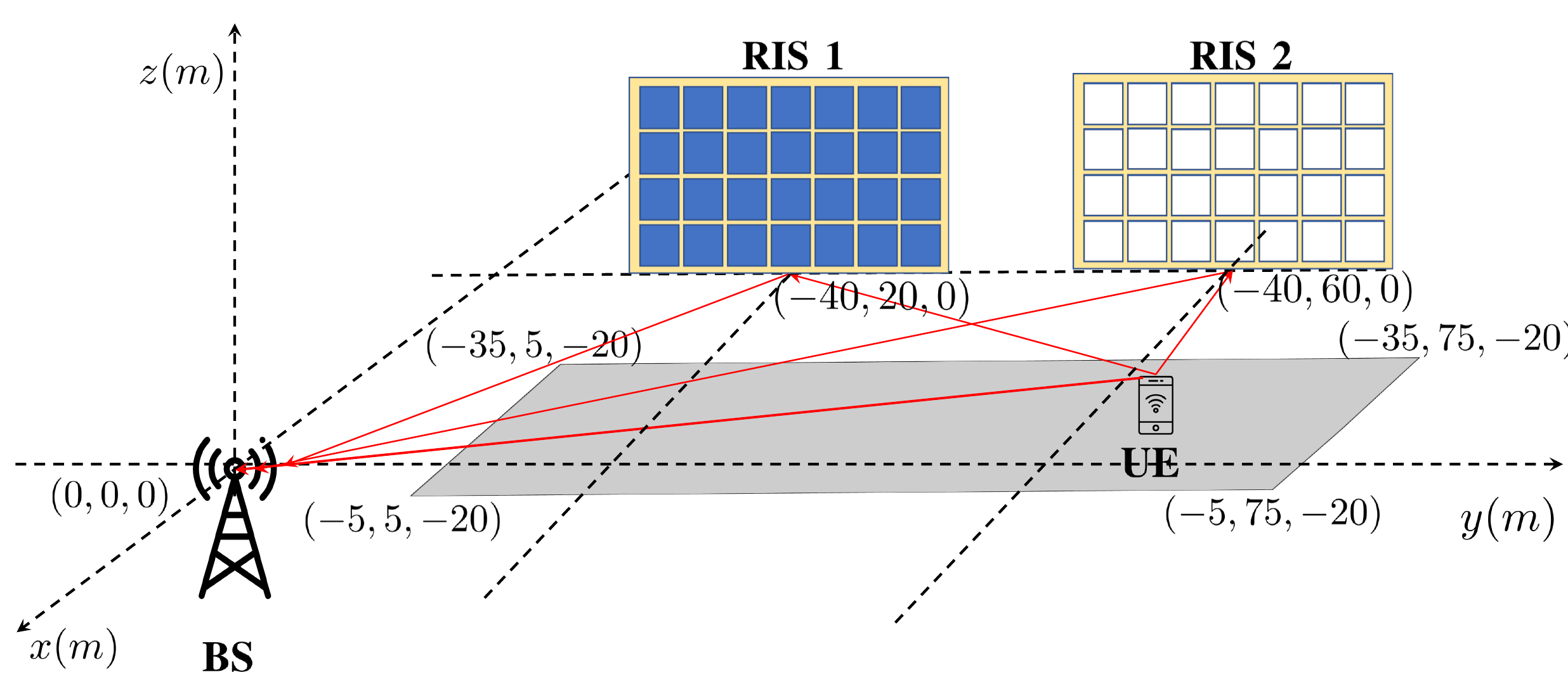}
\caption{Simulation environment for localization in a two-RIS MISO network.} 
\label{label:simsetupmisonetwork}%
\end{figure}

\section{Numerical Results for Multi-RIS MISO Network}
\label{sec.numerical_irs_multi}

In this section, we test the proposed VQ-C network on a setting, where the BS is equipped with multiple antennas and multiple RISs are deployed to aid localization. Here, we need to design both a BS beamforming codebook, and a common RIS codebook, which is shared amongst multiple RISs. 
The design of a shared codebook is inherently more challenging than the design of a codebook for a single RIS, as the shared codebook should contain individually optimized codewords for each RIS and also codewords for multiple RISs that enable constructive interference when deployed together for focusing toward the target user. The codeword selection mechanism also needs to enable the drawing of multiple codewords for multiple RISs in a coordinated fashion in each time frame. 






\subsection{Simulation Environment}

The multi-RIS MISO system is constructed such that a BS, equipped with one RF chain, $M=8$ antennas and $B=500$ codebook size, is placed at $(0m,0m,0m)$. 
The RISs, each with $8 \times 8$ reflective elements and sharing a common RIS codebook with size $V=5000$, are placed at the edge of the network to assist UE localization. 
The $2$-RIS network is depicted in Fig.~\ref{label:simsetupmisonetwork} and the positions of RIS 1 and RIS 2 are $(-40m,20m,0m)$ and $(-40m,60m,0m)$ respectively. 
The RISs in the $3$-RIS network are placed at $(-40m,20m,0m)$, $(-40m,40m,0m)$ and $(-40m,60m,0m)$ respectively. 





The user distribution and channel model adopted here are consistent with the ones in Section \ref{siso_singleris_environment}, where 
the UE location $\bm{p}$ is generated uniformly within a rectangular service area on the $x$-$y$ plane $(-20m\pm15m, 40m\pm35m, -20m)$, as shown in Fig.~\ref{label:simsetupmisonetwork}. 
We set the Rician factor $\epsilon$ to $10$, and set ${2\pi d_{A}}/{\lambda_c} =  {2\pi d_{R}}/{\lambda_c}$ to $1$. 
The path loss models (in dB) of the direct and reflected channel are $32.6+36.7\log(d_1)$ and $30+22\log(d_2)$, respectively, where $d_1$ denotes the distance in the direct path (between UE and BS), and $d_2$ denotes the distance in the reflected path (between BS/$k$-th RIS, between UE/$k$-th RIS)\cite{pathloss}. 
We use $10$ MHz bandwidth with a noise power spectral density of $-170$ dBm/Hz. Again, this is the same simulation setting as in \cite{twcactive}.


%

\begin{figure}
\centering
\includegraphics[width=\columnwidth]{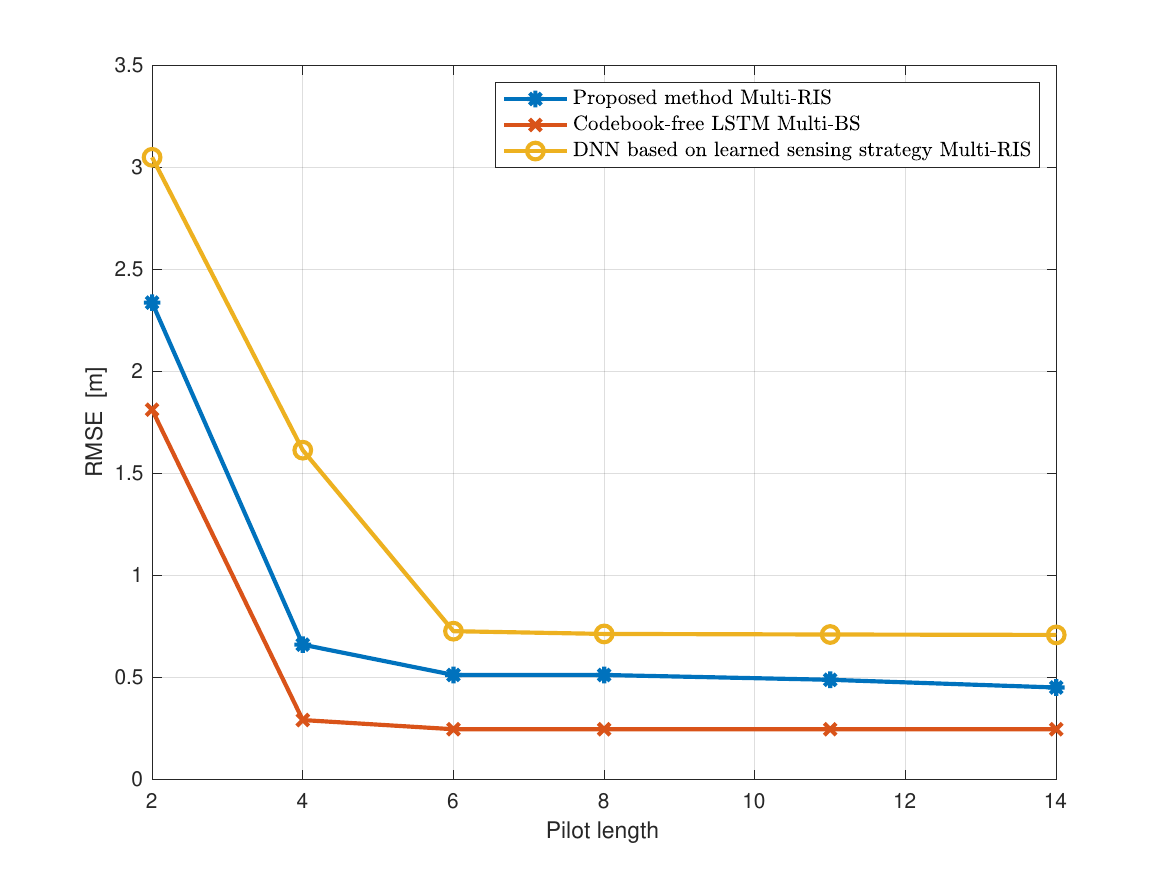}%
\caption{Localization error vs. pilot length with raw SNR$=25$dB, codebook size $V = 5000$, $B = 500$ in a two-RIS MISO network.}%
\label{fig.1BS2RIS}%
\end{figure}

\subsection{Simulation Results}
The localization performance of multi-RIS MISO network with respect to the pilot length is depicted in Fig.~\ref{fig.1BS2RIS}.  
Upon close examination of Fig.~\ref{fig.1BS2RIS}, we observe that the proposed VQ-C active sensing algorithm is seen to have a high localization accuracy, with performance approaching its codebook-free counterpart with $3000$ RIS codewords and $500$ BS beamforming codewords, which amounts to $13$ bits and $9$ bits per control signal to specify each RIS codeword and each BS beamforming codeword respectively. However, the performance gap is also not zero, similar to the phenomenon observed in the single SISO setting from Section \ref{sec.numerical_irs_single},  due to the neural network's limitation in traversing a discrete function space compared to continuous ones. 
This result is also consistent with the conclusion in Section \ref{sec.numerical_irs_single}, which implies that the proposed neural network is capable of constructing BS beamforming codebook and RIS codebook, from which the codewords are sequentially drawn to minimize the localization error based on current and historical measurements.

Lastly, we compare the performance of the multi-RIS-assisted MISO system (as shown in Fig.\ref{fig.1BS2RIS}) with that of the single-RIS-assisted SISO system (as depicted in Fig.\ref{fig.loc_riscodebook_mse_vs_t}). 
By evaluating the proposed active sensing algorithm and the learned sensing algorithm from these figures, we observe an improvement in localization accuracy in the multi-RIS-assisted MISO network over the single-RIS-assisted SISO network. 
This suggests that the additional codewords drawn by the second RIS and the BS improve beamforming capabilities (as opposed to only having one RIS), resulting in enhanced localization accuracy. 

\begin{figure}
\centering
\includegraphics[width=\columnwidth]{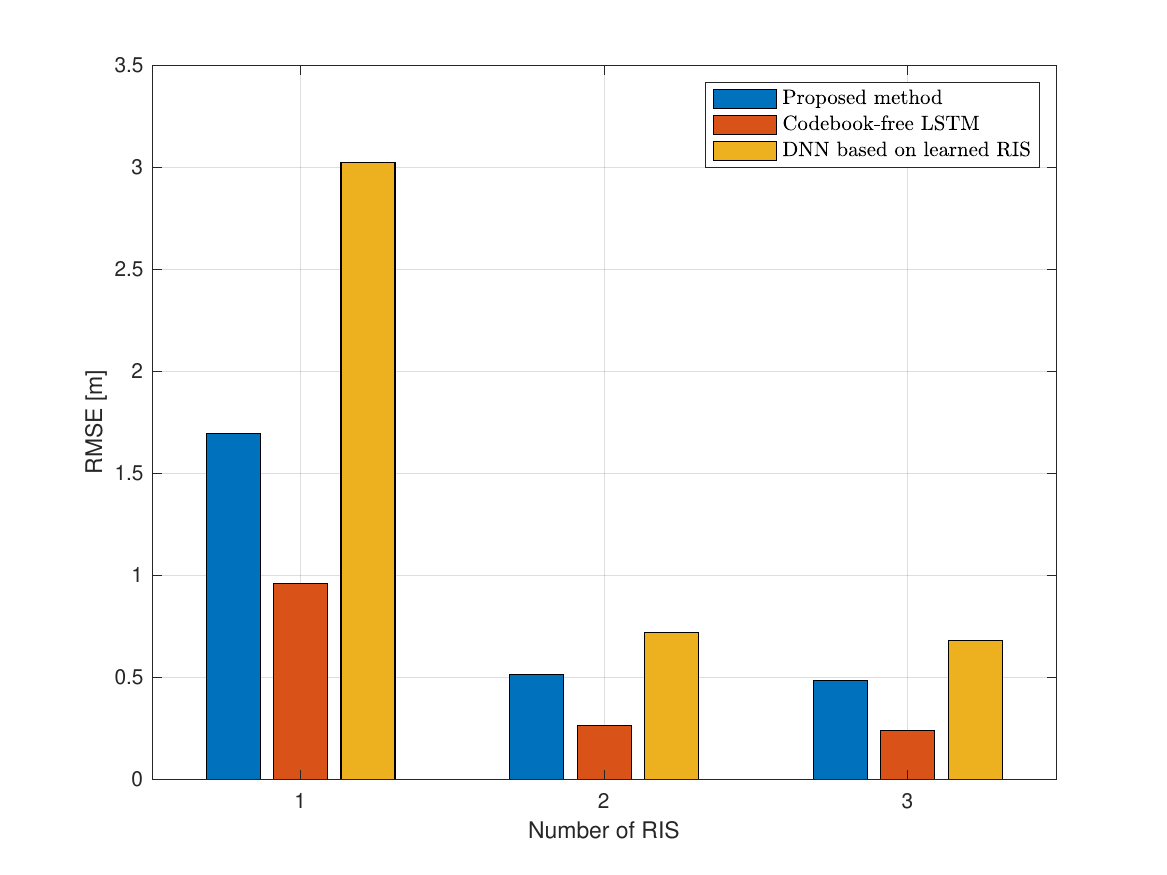}%
\caption{Localization error vs. number of RISs with $T = 6$, raw SNR$=25$dB, codebook size $V = 5000$, $B = 500$ in a MISO network. }%
\label{fig.codebookbar}%
\end{figure}

In terms of the performance under different numbers of RIS, we report here that the learning-based approach is able to work with different numbers of RISs, from 1 RIS to 3 RISs, as depicted in Fig.~\ref{fig.codebookbar}. As the number of RISs increases, the associated localization accuracy increases. The performance gain is due to the extra angular information about the UE provided by the additional RISs.



\subsection{Interpretation of the Solution Learned by VQ-C}

\begin{figure}[!t]
  \centering
  \begin{subfigure}{.33\columnwidth}
    \centering
    \includegraphics[width=\linewidth]{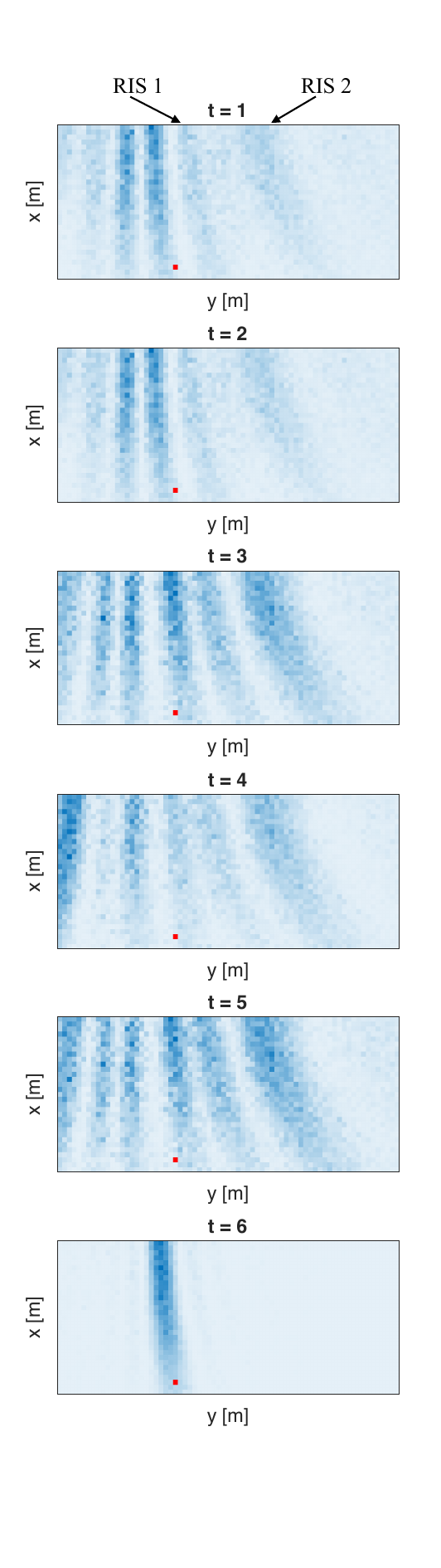}
         \vspace{-3em}
\caption{RIS 1 radio map.}
    \label{loc_ris_interp_6_rnn_ris1}
  \end{subfigure}%
  \hfill
  \begin{subfigure}{.33\columnwidth}
    \centering
    \includegraphics[width=\linewidth]{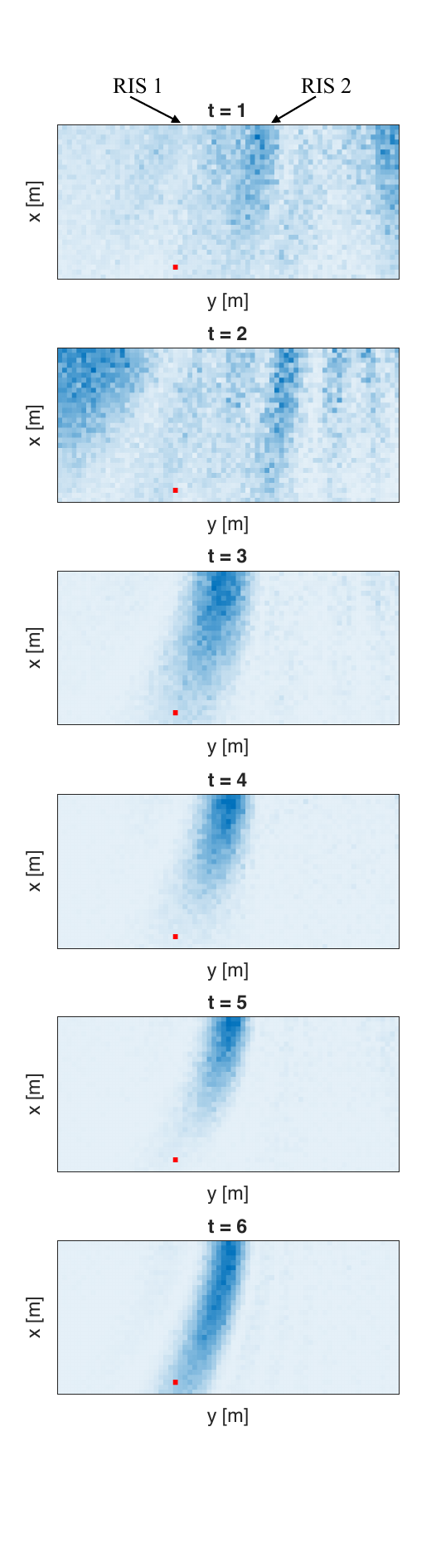}
         \vspace{-3em}
\caption{RIS 2 radio map.}%
    \label{loc_ris_interp_6_rnn_ris2}
  \end{subfigure}%
  \hfill
   \begin{subfigure}{.33\columnwidth}
    \centering
    \includegraphics[width=\linewidth]{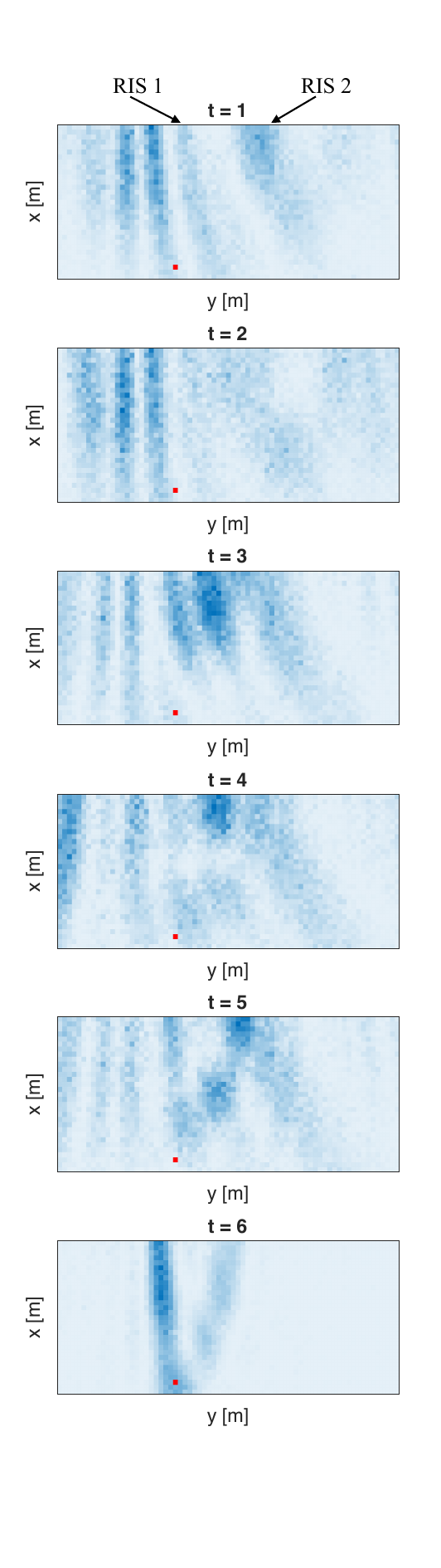}
         \vspace{-3em}
\caption{RIS 1 and 2 radio map.}
    \label{loc_ris_interp_6_rnn_bothris}
  \end{subfigure}%
  \hfill
\caption{RSS distribution of the proposed VQ-C method from time frame $1$ to $6$ with raw SNR = $35$dB in a two-RIS MISO network.}
    \label{loc_2ris_interpretation}
\end{figure}

We interpret the sensing codewords drawn from their respective learned codebooks from the proposed
VQ-C network. 
Here, we test the VQ-C network for a single user with an arbitrary position and use the RSS
distribution (or radio map) as a means to demonstrate the adaptive sequence of beamforming codewords used by the BS and both RISs. 
To achieve this, at each time frame, we separately record the selected codewords from the RIS codebook for RIS 1 and RIS 2, and the selected codeword from the BS beamforming codebook. 
We then plot the three resulting radio maps across the service area over the $x-y$ plane, with RSS measurements taken at every $1m \times 1m$ block from time frame $1$ to $6$, as depicted in Fig.~\ref{loc_2ris_interpretation}. These three radio maps represent: 1) RSS distribution generated by the selected codewords of RIS 1 and BS; 2) RSS distribution generated by the selected codewords of RIS 2 and BS; 3) RSS distribution generated by the selected codewords of RIS 1, RIS 2 and BS together. 
Here, we use the red dot to denote the true position of the UE.
From Fig.\ref{loc_2ris_interpretation}(\subref{loc_ris_interp_6_rnn_ris1}), it is evident that RIS 1 adopts a sequence of codewords that initially probe the environment with broad beams during the first few time frames to search for the UE, then gradually focus toward the UE with narrower beams. A similar strategy is adopted by RIS 2, as shown in Fig.\ref{loc_2ris_interpretation}(\subref{loc_ris_interp_6_rnn_ris2}), which selects a sequence of codewords ranging from broad beams to narrow beams. The combined beampattern resulting from the selected BS and RIS codewords, depicted in Fig.~\ref{loc_2ris_interpretation}(\subref{loc_ris_interp_6_rnn_bothris}), reveals constructive interference near the UE and destructive interference outside this area as more measurements become available. This interference pattern is the key to achieving accurate localization. 
It shows that the proposed VQ-C network is capable of designing meaningful and interpretable RIS codebook and BS beamforming codebook, from which an adaptive sequence of codewords is chosen to narrow down the UE location.

\section{conclusions}
\label{sec.conclusions}

This paper proposes a learning-based BS beamforming and RIS codebook design and a learning-based codeword selection strategy for active sensing in an uplink localization setting. By integrating VQ-VAE and LSTM networks to learn the discrete space of the codebooks and to capture the temporal features, we enable the adaptive selection of BS beamforming codewords and RIS codewords from the respective codebooks based on the sequence of received measurements. Numerical results show that the proposed codebook effectively enables active sensing to achieve low localization errors without exhaustive beam training overhead, with performance approaching its codebook-free counterpart. The proposed solution demonstrates interpretable results in both the single-RIS SISO network and the multi-RIS MISO network.


\bibliography{references}

\begin{thebibliography}{10}
\providecommand{\url}[1]{#1}
\csname url@samestyle\endcsname
\providecommand{\newblock}{\relax}
\providecommand{\bibinfo}[2]{#2}
\providecommand{\BIBentrySTDinterwordspacing}{\spaceskip=0pt\relax}
\providecommand{\BIBentryALTinterwordstretchfactor}{4}
\providecommand{\BIBentryALTinterwordspacing}{\spaceskip=\fontdimen2\font plus
\BIBentryALTinterwordstretchfactor\fontdimen3\font minus
  \fontdimen4\font\relax}
\providecommand{\BIBforeignlanguage}[2]{{%
\expandafter\ifx\csname l@#1\endcsname\relax
\typeout{** WARNING: IEEEtran.bst: No hyphenation pattern has been}%
\typeout{** loaded for the language `#1'. Using the pattern for}%
\typeout{** the default language instead.}%
\else
\language=\csname l@#1\endcsname
\fi
#2}}
\providecommand{\BIBdecl}{\relax}
\BIBdecl

\bibitem{codebookspawc}
Z.~Zhang and W.~Yu, ``Codebook learning for active sensing with reconfigurable
  intelligent surface,'' in \emph{IEEE Int.\ Workshop Signal Process.\ Adv.\
  Wireless Commun.\ (SPAWC)}, Sept. 2024, pp. 821--825.

\bibitem{bible}
Q.~Wu \emph{et~al.}, ``Intelligent surfaces empowered wireless network: Recent
  advances and the road to {6G},'' \emph{Proc. IEEE}, pp. 1--40, 2024.

\bibitem{survey_ris}
Q.~Wu and R.~Zhang, ``Towards smart and reconfigurable environment: Intelligent
  reflecting surface aided wireless network,'' \emph{IEEE Commun.\ Mag.},
  vol.~58, no.~1, pp. 106--112, Nov. 2020.

\bibitem{hguo}
H.~{Guo}, Y.~{Liang}, J.~{Chen}, and E.~G. {Larsson}, ``Weighted sum-rate
  maximization for reconfigurable intelligent surface aided wireless
  networks,'' \emph{IEEE Trans.\ Wireless Commun.}, vol.~19, no.~5, pp.
  3064--3076, May 2020.

\bibitem{nulling}
T.~Jiang and W.~Yu, ``Interference nulling using reconfigurable intelligent
  surface,'' \emph{IEEE J.\ Select.\ Areas Commun.}, vol.~40, no.~5, pp.
  1392--1406, Jan. 2022.

\bibitem{scheduling}
Z.~Zhang, T.~Jiang, and W.~Yu, ``Learning based user scheduling in
  reconfigurable intelligent surface assisted multiuser downlink,'' \emph{IEEE
  J.\ Sel.\ Topics Signal Process.}, vol.~16, no.~5, pp. 1026--1039, May 2022.

\bibitem{codewordsurvey}
J.~An, C.~Xu, Q.~Wu, D.~W.~K. Ng, M.~Di~Renzo, C.~Yuen, and L.~Hanzo,
  ``Codebook-based solutions for reconfigurable intelligent surfaces and their
  open challenges,'' \emph{IEEE Wireless Commun.}, vol.~31, no.~2, pp.
  134--141, Apr. 2024.

\bibitem{randomcodebook}
J.~An and L.~Gan, ``The low-complexity design and optimal training overhead for
  {IRS}-assisted {MISO} systems,'' \emph{IEEE Wireless Commun.\ Lett.},
  vol.~10, no.~8, pp. 1820--1824, May 2021.

\bibitem{codebook_euclideanmax}
J.~An, C.~Xu, L.~Gan, and L.~Hanzo, ``Low-complexity channel estimation and
  passive beamforming for {RIS}-assisted {MIMO} systems relying on discrete
  phase shifts,'' \emph{IEEE Trans.\ Commun.}, vol.~70, no.~2, pp. 1245--1260,
  Nov. 2022.

\bibitem{codebook_euclideanmax2}
J.~An, C.~Xu, L.~Wang, Y.~Liu, L.~Gan, and L.~Hanzo, ``Joint training of the
  superimposed direct and reflected links in reconfigurable intelligent surface
  assisted multiuser communications,'' \emph{IEEE Trans.\ Green Commun.\
  Netw.}, vol.~6, no.~2, pp. 739--754, Jan. 2022.

\bibitem{fieldtest}
X.~Pei, H.~Yin, L.~Tan, L.~Cao, Z.~Li, K.~Wang, K.~Zhang, and E.~Björnson,
  ``{RIS}-aided wireless communications: Prototyping, adaptive beamforming, and
  indoor/outdoor field trials,'' \emph{IEEE Trans.\ Commun.}, vol.~69, no.~12,
  pp. 8627--8640, Sept. 2021.

\bibitem{ringtype}
F.~Wang, X.~Wang, X.~Li, X.~Hou, L.~Chen, S.~Suyama, and T.~Asai, ``Ring-type
  codebook design for reconfigurable intelligent surface near-field
  beamforming,'' in \emph{Proc.\ IEEE Int.\ Symp.\ Pers.\ Indoor Mobile Radio
  Commun.\ (PIMRC)}, Sept. 2022, pp. 391--396.

\bibitem{env_aware}
Z.~Yu, J.~An, E.~Basar, L.~Gan, and C.~Yuen, ``Environment-aware codebook
  design for {RIS}-assisted {MU}-{MISO} communications: Implementation and
  performance analysis,'' \emph{IEEE Trans.\ Commun.}, vol.~72, no.~12, pp.
  7466--7479, Jun. 2024.

\bibitem{codebook_learning_RL}
\BIBentryALTinterwordspacing
Y.~Zhang and A.~Alkhateeb, ``Learning reflection beamforming codebooks for
  arbitrary {RIS} and non-stationary channels,'' 2021. [Online]. Available:
  \url{https://arxiv.org/abs/2109.14909}
\BIBentrySTDinterwordspacing

\bibitem{codebook_RL_3}
A.~Abdallah, A.~Celik, M.~M. Mansour, and A.~M. Eltawil, ``Deep reinforcement
  learning based beamforming codebook design for {RIS}-aided {mmWave}
  systems,'' in \emph{IEEE Consumer Commun.\ Netw.\ Conf.\ (CCNC)}, Mar. 2023,
  pp. 1020--1026.

\bibitem{codebook_learning_DNN}
J.~Kim, S.~Hosseinalipour, A.~C. Marcum, T.~Kim, D.~J. Love, and C.~G. Brinton,
  ``Learning-based adaptive {IRS} control with limited feedback codebooks,''
  \emph{IEEE Trans.\ Wireless Commun.}, vol.~21, no.~11, pp. 9566--9581, Jun.
  2022.

\bibitem{hejiguang}
J.~He, H.~Wymeersch, T.~Sanguanpuak, O.~Silven, and M.~Juntti, ``Adaptive
  beamforming design for {mmWave} {RIS}-aided joint localization and
  communication,'' in \emph{Proc. IEEE Wireless Commun. Netw. Conf. Workshops
  (WCNC-W)}, Apr. 2020, pp. 1--6.

\bibitem{hierarchical1}
X.~Wei, L.~Dai, Y.~Zhao, G.~Yu, and X.~Duan, ``Codebook design and beam
  training for extremely large-scale {RIS}: Far-field or near-field?''
  \emph{China Commun.}, vol.~19, no.~6, pp. 193--204, Jun. 2022.

\bibitem{hiarchicalwuqq}
X.~Liu, Q.~Wu, D.~Hu, R.~Wang, and J.~Wu, ``Hierarchical codebook design and
  analytical beamforming solution for {IRS} assisted communication,''
  \emph{IEEE Trans.\ Wireless Commun.}, vol.~23, no.~8, pp. 8924--8938, Jan.
  2024.

\bibitem{hierarchical2layer}
\BIBentryALTinterwordspacing
T.~Wang, J.~Lv, H.~Tong, C.~You, and C.~Yin, ``A novel two-layer codebook based
  near-field beam training for intelligent reflecting surface,'' 2023.
  [Online]. Available: \url{https://arxiv.org/abs/2303.06962}
\BIBentrySTDinterwordspacing

\bibitem{active2023icc}
Z.~Zhang, T.~Jiang, and W.~Yu, ``Active sensing for localization with
  reconfigurable intelligent surface,'' in \emph{Proc. IEEE Int. Conf. Commun.
  (ICC)}, Jun. 2023, pp. 4261--4266.

\bibitem{twcactive}
------, ``Localization with reconfigurable intelligent surface: An active
  sensing approach,'' \emph{IEEE Trans.\ Wireless Commun.}, vol.~23, no.~7, pp.
  7698--7711, Jul. 2024.

\bibitem{sohrabi2021active}
F.~Sohrabi, T.~Jiang, W.~Cui, and W.~Yu, ``Active sensing for communications by
  learning,'' \emph{IEEE J.\ Select.\ Areas Commun.}, vol.~40, no.~6, pp.
  1780--1794, Jun. 2022.

\bibitem{discretereplearning}
A.~van~den Oord, O.~Vinyals, and K.~Kavukcuoglu, ``Neural discrete
  representation learning,'' in \emph{Proc. Int. Conf. Neural Inf. Process.
  Syst.}, Dec. 2017, p. 6309–6318.

\bibitem{LSTM}
S.~Hochreiter and J.~Schmidhuber, ``Long short-term memory,'' \emph{Neural
  Comput.}, vol.~9, no.~8, p. 1735–1780, Nov. 1997.

\bibitem{shi2022vqtrnntransducersusing}
J.~Shi, G.~Saon, D.~Haws, S.~Watanabe, and B.~Kingsbury, ``{VQ-T}: {RNN}
  transducers using vector-quantized prediction network states,'' in
  \emph{Proc. Interspeech}, 2022, p. 1656–1660.

\bibitem{GRU}
A.~Sant, A.~Abdi, and J.~Soriaga, ``Deep sequential beamformer learning for
  multipath channels in mmwave communication systems,'' in \emph{Proc.\ IEEE
  Int.\ Conf.\ Acoust.\ Speech, Signal Process. (ICASSP)}, May 2022, pp.
  5198--5202.

\bibitem{pingpong}
T.~Jiang, F.~Sohrabi, and W.~Yu, ``Active sensing for two-sided beam alignment
  and reflection design using ping-pong pilots,'' \emph{IEEE J.\ Sel.\ Areas
  Inf.\ Theory}, vol.~4, pp. 24--39, May 2023.

\bibitem{taomimo}
T.~Jiang and W.~Yu, ``Active sensing for reciprocal {MIMO} channels,''
  \emph{IEEE Trans.\ Signal Process.}, vol.~72, pp. 2905--2920, Jun. 2024.

\bibitem{hantracking}
H.~Han, T.~Jiang, and W.~Yu, ``Active sensing for multiuser beam tracking with
  reconfigurable intelligent surface,'' \emph{IEEE Trans.\ Wireless Commun.},
  vol.~24, no.~1, pp. 540--554, Jan. 2025.

\bibitem{yinghanactive}
Y.~Li and W.~Yu, ``Localization in multipath environments via active sensing
  with reconfigurable intelligent surfaces,'' \emph{IEEE Commun.\ Lett.},
  vol.~28, no.~9, pp. 2061--2065, Jul. 2024.

\bibitem{compress}
X.~Yu, D.~Li, Y.~Xu, and Y.-C. Liang, ``Convolutional autoencoder-based phase
  shift feedback compression for intelligent reflecting surface-assisted
  wireless systems,'' \emph{IEEE Commun.\ Lett.}, vol.~26, no.~1, pp. 89--93,
  Oct. 2022.

\bibitem{lightweight}
X.~Yu and D.~Li, ``Phase shift compression for control signaling reduction in
  {IRS}-aided wireless systems: Global attention and lightweight design,''
  \emph{IEEE Trans.\ Wireless Commun.}, vol.~23, no.~8, pp. 8528--8541, Aug.
  2024.

\bibitem{RNN}
D.~E. Rumelhart and J.~L. McClelland, Eds., \emph{Parallel Distributed
  Processing: Explorations in the Microstructure of Cognition}.\hskip 1em plus
  0.5em minus 0.4em\relax MIT Press, 1987, ch.8, pp.318-362.

\bibitem{taojournal}
T.~{Jiang}, H.~V. {Cheng}, and W.~{Yu}, ``Learning to reflect and to beamform
  for intelligent reflecting surface with implicit channel estimation,''
  \emph{IEEE J.\ Select.\ Areas Commun.}, vol.~39, no.~7, pp. 1931--1945, Jul.
  2021.

\bibitem{multiAIRS}
T.~Zhou, K.~Xu, Z.~Shen, W.~Xie, D.~Zhang, and J.~Xu, ``{AoA}-based positioning
  for aerial intelligent reflecting surface-aided wireless communications: An
  angle-domain approach,'' \emph{IEEE Commun.\ Lett.}, vol.~11, no.~4, pp.
  761--765, Apr. 2022.

\bibitem{pathloss}
Q.~Wu and R.~Zhang, ``Intelligent reflecting surface enhanced wireless network
  via joint active and passive beamforming,'' \emph{IEEE Trans.\ Wireless
  Commun.}, vol.~18, no.~11, pp. 5394--5409, Aug. 2019.

\bibitem{tensorflow}
M.~{Abadi} \emph{et~al.}, ``Tensorflow: A system for large-scale machine
  learning,'' in \emph{Proc. USENIX Conf. Operating Syst. Des. Implementation
  (OSDI)}, 2016, p. 265–283.

\bibitem{adam}
\BIBentryALTinterwordspacing
D.~P. Kingma and J.~Ba, ``Adam: {A} method for stochastic optimization,'' in
  \emph{Proc. Int. Conf. Learn. Representations (ICLR)}, 2015. [Online].
  Available: \url{http://arxiv.org/abs/1412.6980}
\BIBentrySTDinterwordspacing

\bibitem{crlbbound2}
Y.~Noam and H.~Messer, ``Notes on the tightness of the hybrid {C}ramér–{Rao}
  lower bound,'' \emph{IEEE Trans.\ Signal Process.}, vol.~57, no.~6, pp.
  2074--2084, Feb. 2009.

\end{thebibliography}
\bibliographystyle{IEEEtran}

\end{document}